\documentclass[superscriptaddress, aps, showpacs, floatfix, prb,twocolumn,a4paper]{revtex4-1}

\usepackage{graphicx}   

\usepackage{dcolumn}

\usepackage{hyperref}
\usepackage{bm}                 
\usepackage{amsmath}
\usepackage{amsfonts}
\usepackage{revsymb}
\usepackage{color}

\def\wn/{\,cm$^{-1}$}
\def\area/{\,cm$^{-2}$}
\def\cubic/{$_\mathrm{c}$}
\def\DM/{Dzyaloshinskii-Moriya}

\def\bcgo/{Ba$_2$CoGe$_2$O$_7$}
\def\scso/{Sr$_2$CoSi$_2$O$_7$}
\def\ccso/{Ca$_2$CoSi$_2$O$_7$}
\def\nfbo/{NdFe$_3$(BO$_3$)$_4$}
\def\tfbo/{TbFe$_3$(BO$_3$)$_4$}
\def\bfo/{BiFeO$_3$}
\def\azurite/{Cu$_3$(CO$_3$)$_2$(OH)$_2$}
\def\lco/{LiCu$_2$O$_2$}
\def\tmo/{TbMnO$_3$}

\def\edc/{\ensuremath{\mathbf{E}}}
\def\eac/{\ensuremath{\mathbf{E}_\omega}}
\def\bdc/{\ensuremath{\mathbf{B}}}
\def\hdc/{\ensuremath{\mathbf{H}}}
\def\bac/{\ensuremath{\mathbf{B}_\omega}}
\def\hac/{\ensuremath{\mathbf{H}_\omega}}

\def\kvec/{\ensuremath{\mathbf{k}}}
\def\Pvec/{\ensuremath{\mathbf{P}}}
\def\Mvec/{\ensuremath{\mathbf{M}}}
\def\Tvec/{\ensuremath{\mathbf{T}}}
\def\Svec/{\ensuremath{\mathbf{S}}}

\newcommand{\vect}[1]{\ensuremath{{\bf #1}}}

\def\ket#1{\left| #1 \right>}
\def\bra#1{\left< #1 \right|}

\begin{document}

\title{Directional dichroism in the paramagnetic state of multiferroics: a case study of infrared light absorption in \scso/ at high temperatures}

\author{J. Viirok}
\author{U. Nagel}
\author{T. R{\~o}{\~o}m}
\affiliation{National Institute of Chemical Physics and Biophysics, Akadeemia tee 23, 12618 Tallinn, Estonia}

\author{D. Farkas}
\affiliation{Department of Physics, Budapest University of Technology and Economics and MTA-BME Lend\"ulet Magneto-optical Spectroscopy Research Group, 1111 Budapest, Hungary}

\author{P. Balla}
\affiliation{Institute for Solid State Physics and Optics, Wigner Research Centre for Physics, Hungarian Academy of Sciences, PO Box. 49, H-1525 Budapest, Hungary}

\author{D. Szaller}
\affiliation{Department of Physics, Budapest University of Technology and Economics and MTA-BME Lend\"ulet Magneto-optical Spectroscopy Research Group, 1111 Budapest, Hungary}
\affiliation{Institute of Solid State Physics, Vienna University of Technology, 1040 Vienna, Austria}

\author{V. Kocsis}
\affiliation{Department of Physics, Budapest University of Technology and Economics and MTA-BME Lend\"ulet Magneto-optical Spectroscopy Research Group, 1111 Budapest, Hungary}
\affiliation{RIKEN Center for Emergent Matter Science (CEMS), Wako 351-0198, Japan}

\author{Y. Tokunaga}
\affiliation{RIKEN Center for Emergent Matter Science (CEMS), Wako, Saitama 351-0198, Japan}
\affiliation{Department of Advanced Materials Science, University of Tokyo, Kashiwa 277-8561, Japan}

\author{Y. Taguchi}
\affiliation{RIKEN Center for Emergent Matter Science (CEMS), Wako 351-0198, Japan}

\author{Y. Tokura}
\affiliation{RIKEN Center for Emergent Matter Science (CEMS), Wako, Saitama 351-0198, Japan}
\affiliation{Department of Applied Physics, University of Tokyo, Hongo, Tokyo 113-8656, Japan}

\author{B. Bern\'ath}
\author{D. L. Kamenskyi}
\affiliation{High Field Magnet Laboratory (HFML-EMFL), Radboud University, Toernooiveld 7, 6525 ED Nijmegen, The Netherlands}

\author{I. K{\'e}zsm{\'a}rki}
\affiliation{Department of Physics, Budapest University of Technology and Economics and MTA-BME Lend\"ulet Magneto-optical Spectroscopy Research Group, 1111 Budapest, Hungary}
\affiliation{Experimental Physics V, Center for Electronic Correlations and Magnetism, Institute of Physics, University of Augsburg, 86159 Augsburg, Germany}

\author{S. Bord{\'a}cs}
\affiliation{Department of Physics, Budapest University of Technology and Economics and MTA-BME Lend\"ulet Magneto-optical Spectroscopy Research Group, 1111 Budapest, Hungary}
\affiliation{Hungarian Academy of Sciences, Premium Postdoctor Program, 1051 Budapest, Hungary}

\author{K. Penc}
\affiliation{Department of Physics, Budapest University of Technology and Economics and MTA-BME Lend\"ulet Magneto-optical Spectroscopy Research Group, 1111 Budapest, Hungary}
\affiliation{Institute for Solid State Physics and Optics, Wigner Research Centre for Physics, Hungarian Academy of Sciences, PO Box. 49, H-1525 Budapest, Hungary}

\date{\today}

\begin{abstract}
The coexisting magnetic and ferroelectric orders in multiferroic materials give rise to a handful of novel magnetoelectric phenomena, such as the absorption difference for the opposite propagation directions of light called the non-reciprocal directional dichroism (NDD). Usually these effects are restricted to low temperature, where the multiferroic phase develops. In this paper we report the observation of NDD in the paramagnetic phase of \scso/ up to temperatures more than ten times higher than its N\'eel temperature (7\,K) and in fields up to 30\,T. The magnetically induced polarization and NDD in the disordered paramagnetic phase is readily explained by the single-ion spin-dependent hybridization mechanism, which does not necessitate correlation effects between magnetic ions. 
The \scso/ provides an ideal system for a theoretical case study, demonstrating the concept of magnetoelectric spin excitations in a paramagnet via analytical as well as numerical approaches. We applied exact diagonalization of a spin cluster to map out the temperature and field dependence of the spin excitations, as well as symmetry arguments of the single ion and lattice problem to get the spectrum and  selection rules.
\end{abstract}

\keywords{multiferroics, directional dichroism, paramagnetic, \scso}

\maketitle

\section{Introduction}
Non-reciprocal directional dichroism (NDD) is the property of a material to have different absorption coefficients for light propagation directions, $\pm\kvec/$ along and opposite to a given direction in the crystal.\cite{BarronBook2009} Although NDD was observed long time ago for the exciton transitions of the polar semiconductor CdS,\cite{Hopfield1960} it was recognized as a general magneto-optical phenomenon of non-centrosymmetric materials only by the seminal works of Rikken and his co-workers.\cite{Rikken1997,Rikken2002}
The two basic cases of NDD were identified as the {\it magneto-chiral dichroism}\cite{Rikken1997} (MChD)  and the {\it toroidal dichroism}. In the case of MChD the absorption coefficient is different for light propagation along and opposite the magnetization of a chiral magnet $\kvec/\parallel\Mvec/$.\cite{Rikken2002} In the case of toroidal dichroism  NDD appears for propagation along and opposite $\kvec/\parallel\Pvec/\times\Mvec/$, where and $\Pvec/$ is the ferroelectric polarization of the material. 
In general, NDD can be finite only when both the spatial inversion and time reversal symmetries are broken, as these symmetry operations interconnect  the light beams propagating in opposite directions. 
Following the same principle, a recent study rigorously classified the magnetic point groups compatible with NDD.\cite{Szaller2013} It predicted that beside magneto-chiral and toroidal dichroism NDD can arise in previously unclassified cases, which cannot be classified by a static vector quantity, such as the magnetization or the toroidal moment.

Magnetoelectric (ME) multiferroics with coexisting ferroelectric and magnetic orders naturally have the low-symmetry ground states exhibiting magneto-chiral or toroidal dichroism, and indeed gigantic NDD was found in their collective excitations, typically  in the GHz-THz range. \cite{Kezsmarki2011, Bordacs2012, Takahashi2012, Takahashi2013, Kezsmarki2014, Kuzmenko2015, Kezsmarki2015}
At these low frequencies, where the electromagnetic radiation is uniform on the scale of the magnetic unit cell,  NDD can solely originate from the coupled dynamics of the magnetization and the electric polarization, the dynamic ME effect.
When light beams travel in a ME material, for a beam propagating in one direction the oscillating magnetization generated by the electric field of the light can enhance the conventional magnetization component, induced by the magnetic field of light, whereas these two terms interfere destructively for the counter-propagating beam since the relative phase of the electric and magnetic fields of light changes by $\pi$ when reversing the propagation direction. The direct connection between the NDD and the ME effect allows the calculation of the direct current (dc) ME coefficient from the NDD spectrum using the sum rule.\cite{Szaller2014}
Thus, the spectroscopic information about the ME resonances can promote the synthesis of new multiferroics with large ME coefficient.

One-way transparency is an extreme case of gigantic NDD when the light is absorbed only in one direction but not  for the opposite propagation direction.
Since no physical law prohibits the one-way transparency,  an efficient one-way light guide can be realized. 
Furthermore, the transparent direction can be switched with applied magnetic\cite{Kezsmarki2011} and possibly with electric field.\cite{Kuzmenko2018,Kocsis2018} 
Some of the multiferroics, e.g.  melilites \bcgo/, \ccso/ and \scso/, are not far from the ideal realization of one-way transparency.\cite{Kezsmarki2014}
In principle these materials can find applications in photonics as diodes for THz radiation or directional light switches.\cite{Kezsmarki2015} 
However, the multiferroic order possessing large NDD usually develops well below room temperature in the known materials, with the exception of the room temperature multiferroic \bfo/ ,\cite{Kezsmarki2015} rendering their use in device applications impractical.

\begin{figure}[tb]
	\includegraphics[width=0.9\columnwidth]{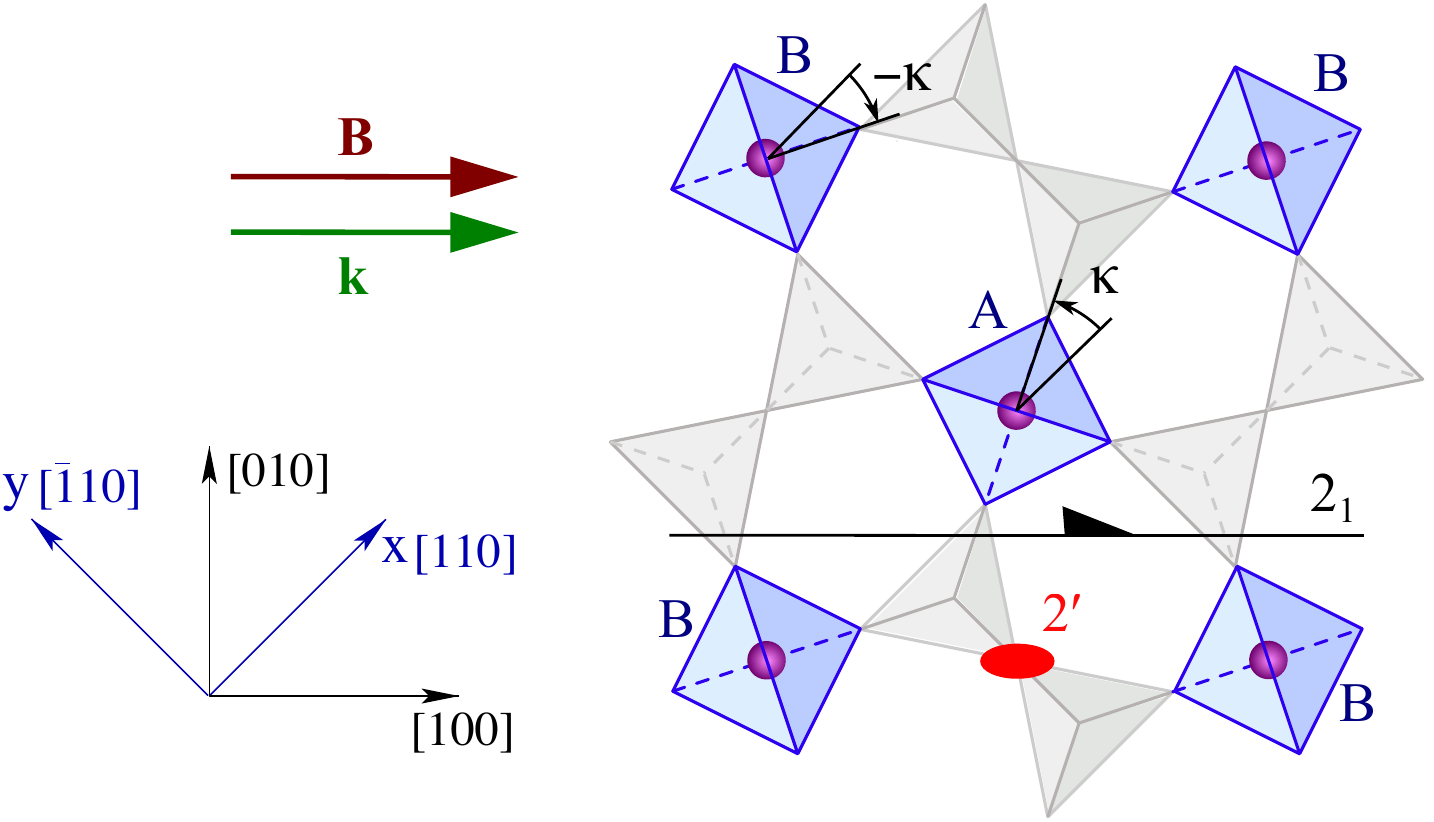}
	\caption{ \label{fig:2sublatt} 
		Schematic illustration of the \protect{\scso/} crystal structure, space group $P\overline{4}2_1m$, and  the coordinate axes, $\vect{x}\parallel [110]$ and $ \vect{y}\parallel [\bar{1}10]$.
		Purple circles denote the $S=3/2$ Co$^{2+}$ ions in the centers of the O$^{2-}$ tetrahedra (shown in blue). The compound is built up by layers of  CoO$_4$ tetrahedra  and Si$_2$O$_7$ units (grey), which are separated by Sr ions. For clarity, only a single layer is shown. $\pm\kappa$ is the rotation angle of tetrahedra about the $[001]$ axis away from $[110]$. A and B denote the two types of tetrahedra with $+\kappa$ and $-\kappa$ tilt. In the experiment the THz light propagates along the direction of the magnetic field, $\kvec/\!\parallel\!\bdc/\!\parallel\! [100]$.  
		The magnetic field breaks the time reversal symmetry and for this field direction the remaining unitary symmetry operation is the $2_1$ screw axis, (black  half-arrow) and the twofold rotation about the $[001]$ axis followed by  time reversal will compose the $2'$ antiunitary operation (red ellipse). 
	 }
\end{figure}

Here, we suggest an alternative way of achieving NDD by demonstrating that the spin excitations in \scso/ shows  NDD well above the anti-ferromagnetic ordering temperature,  $T_\mathrm{N}$=7\,K. 
The  crystal structure of melilites (see Fig.~\ref{fig:2sublatt}), lacks the inversion symmetry. Applying magnetic field  the time reversal symmetry will be broken in the paramagnetic phase and the necessary conditions for NDD are then fulfilled.
Akaki et al.~\cite{Akaki2012, Akaki2013} demonstrated that in \scso/ the dc ME effect persists in the paramagnetic regime, where the field-induced polarization scales with the square of the magnetization. 
Since the  dc ME susceptibility is related to NDD  by the ME sum rule,\cite{Szaller2014} we expect that NDD appears also in the paramagnetic phase of \scso/.

In this paper, motivated by the discovery of the high temperature dc ME effect in  \scso/, we study the NDD  in the paramagnetic phase.
We measured the THz absorption spectra in magnetic fields up to 30\,T over a broad range of temperatures below and well above $T_\mathrm{N}$. 
The magnetic field was applied along the [100] direction, which induces  magnetization parallel to the field. 
The point group of the  \scso/, $\overline{4}2m1'$ in the paramagnetic state, is then reduced to the magnetic point group $22'2'$ .\cite{Bordacs2012} 
In this chiral symmetry MChD is expected to emerge for light propagation along the magnetic field, i.e. in the  Faraday geometry. 
Indeed, our experiments show that the spectra are markedly different in positive and negative magnetic field, which is the hallmark of MChD. 
Using exact diagonalization we reproduced the magnetic field and temperature dependence of the MChD signal. 
To interpret the numerical results, a single-site analytic model was developed which shows that the finite NDD arises  if all three are present,   magnetic field, spin anisotropy and ME coupling.

Recently S. Yu {\it et al.}  studied  spin excitations  of a polar ferrimagnet FeZnMo$_3$O$_8$ in Ref.~[\onlinecite{Yu2018}]. 
They demonstrated that the toroidal dichroism can be realized in the paramagnetic phase when the light propagates along the cross product of the built-in  polarization and the external magnetic field induced magnetization, $\vect{k}\parallel \vect{P}\times \vect{M}$. 
In contrast, here we study MChD, a different form of NDD, and develop  microscopic spin models to understand  NDD in the paramagnetic phase.

The paper is organized as follows.
After description of the experimental methods in Section\,\ref{sec:experimental_methods}, the experimental results are presented in  Section\,\ref{sec:experimental_results}. 
To understand the observed spin excitations and NDD, firstly, a  spin Hamiltonian and spin-induced polarization of \scso/ are introduced in Section\,\ref{sec:ED_spin_ham}.
The Hamiltonian  is numerically diagonalized for a small cluster and the eigenstates are used to calculate magnetic and ME susceptibilities in Section\,\ref{sec:ED}. 
Secondly, using a single-ion model in Section\,\ref{sec:single_ion} the results of the exact diagonalization are interpreted. 
Thirdly, the selection rules found for the single ion case are generalized for the lattice model in Section\,\ref{sec:lattice}. 
In addition, in this section  the effects of the exchange interaction in leading order of perturbation theory are analyzed. 
Finally, the main experimental and theoretical results are summarized in Section\,\ref{sec:summary}.

\section{Experimental Methods}\label{sec:experimental_methods}

The \scso/ crystals were grown by the floating zone method. 
First, SrCO$_3$, Co$_3$O$_4$, and de-hydrated SiO$_2$ were mixed in stoichiometric amount and sintered for 120\,hours at 1200\,$^\circ$C in air with one intermediate re-grinding. 
The resulting product was pressed into a rod shape and re-sintered for 60\,hours under the same conditions as before. 
The polycrystalline rod was melted into a single crystal ingot in a halogen-incandescent lamp floating zone apparatus (SC-N35HD, NEC).

Samples for THz spectroscopy were disk-shaped  single crystals with a diameter of 4\,mm  and thicknesses of $d=0.2$ and 0.5\,mm  in the [100] direction. 
The external magnetic field \bdc/ and the light propagation $\kvec/$ were both in the [100] direction, $\kvec/\!\parallel\! \bdc/\!\parallel\! [100]$. 
The THz radiation was polarized by a wire-grid polarizer deposited on a dielectric film placed a few millimeters away from  the sample in the incident light beam. 
The transmitted intensity was measured with a Martin-Puplett type interferometer (SPS200, Sciencetech, Inc., Ontario,Canada) and 0.3\,K composite silicon bolometer (Infrared Laboratories) in magnetic fields up to 17\,T for both positive and negative fields in the temperature range between 3\,K and 100\,K. 
Measurements above 17\,T were performed with a Genzel-type interferometer (Bruker 113v) and a 1.6\,K composite Si bolometer (Infrared Laboratories).

The absorption coefficient $\alpha^\pm$ for magnetic fields $\pm B$ was calculated as 
\begin{equation}
  \alpha^{\pm}=-\frac{1}{d}\ln \frac{I(\pm B, T)}{I(0\,\mathrm{T}, T_{\mathrm{ref}})}, 
\end{equation}
where $I(\pm B, T)$ is the transmitted intensity in magnetic field $B$ at temperature $T$ and $I(0\,\mathrm{T}, T_{\mathrm{ref}})$ is the transmitted intensity in zero magnetic field  at temperature $T_{\mathrm{ref}}$. 
For the  magnetic field ratios, $T=T_{\mathrm{ref}}$, and the relative absorption spectra are noted as $\alpha^\pm_\mathrm B$.
If $B$ is constant and $T$ is varied, the ratio is denoted by $\alpha^\pm_T$.
$\alpha^\pm$ was determined for two polarizations of THz radiation, $\eac/\parallel [010]$  and $\eac/\parallel [001]$.

The NDD was detected by changing the direction of the magnetic field from $+B$ to $-B$, i.e. from $\bdc/\uparrow\uparrow \kvec/$ to $\bdc/\downarrow\uparrow \kvec/$. Due to the two-fold rotation symmetry along the [001] axis the reversal of \bdc/  is equivalent to the reversal of \kvec/.

\section{Experimental Results}\label{sec:experimental_results}

\begin{figure*}
	\includegraphics[width=0.95\textwidth]{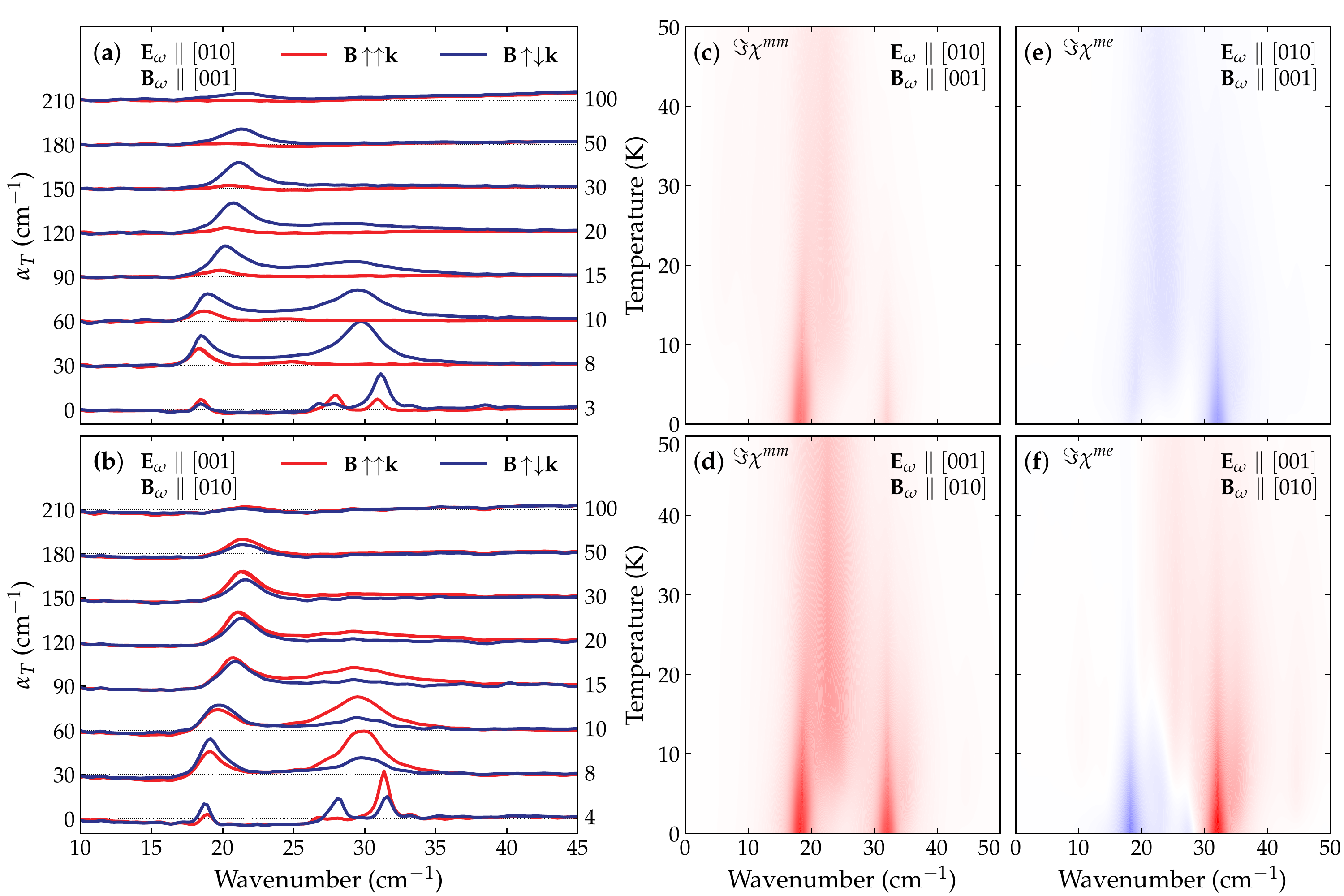}
	\caption{\label{fig:scso_b[100]_14t_experiment}
		Temperature dependence of measured THz absorption spectra (a-b) and calculated susceptibilities (c-f) of \scso/ in 14\,T.
		The THz absorption spectra for magnetic field $ \bdc/\uparrow\uparrow \mathbf{k}\parallel[100]$ (red) and $\bdc/\uparrow\downarrow \mathbf{k}$ (blue) are measured using linearly polarized radiation where in 
		(a)   $\eac/\|[010]$ and $\bac/\|[001]$, and in  (b) the polarization is rotated by $\pi/2$, so that $\eac/\|[001]$ and $\bac/\|[010]$.
		The spectra measured at each temperature are shifted by a constant baseline. 
		(c) and (e) are  the magnetic susceptibility $\Im\chi^{mm}(\omega)$  and the ME susceptibility $\Im\chi^{me}(\omega)$ for the polarization $\eac/\|[010]$ and $\bac/\|[001]$.
		(d) and (f) are   $\Im\chi^{mm}(\omega)$  and  $\Im\chi^{me}(\omega)$ for the polarization $\eac/\|[001]$ and $\bac/\|[010]$. 
		Red (positive) and blue (negative) colors indicate the sign of the susceptibility. 
		The saturation of the color corresponds to the magnitude of the corresponding susceptibility matrix elements, $\Im \chi^{mm}(\omega)$, Eq.\,(\ref{eq:imchimm}), and $\Im \chi^{me}(\omega)$, Eq.\,(\ref{eq:imchime}). 
		The susceptibilities were calculated  by the exact diagonalization of a 4-site cluster.
		}
\end{figure*}

\begin{figure}
	\includegraphics[width=0.45\textwidth]{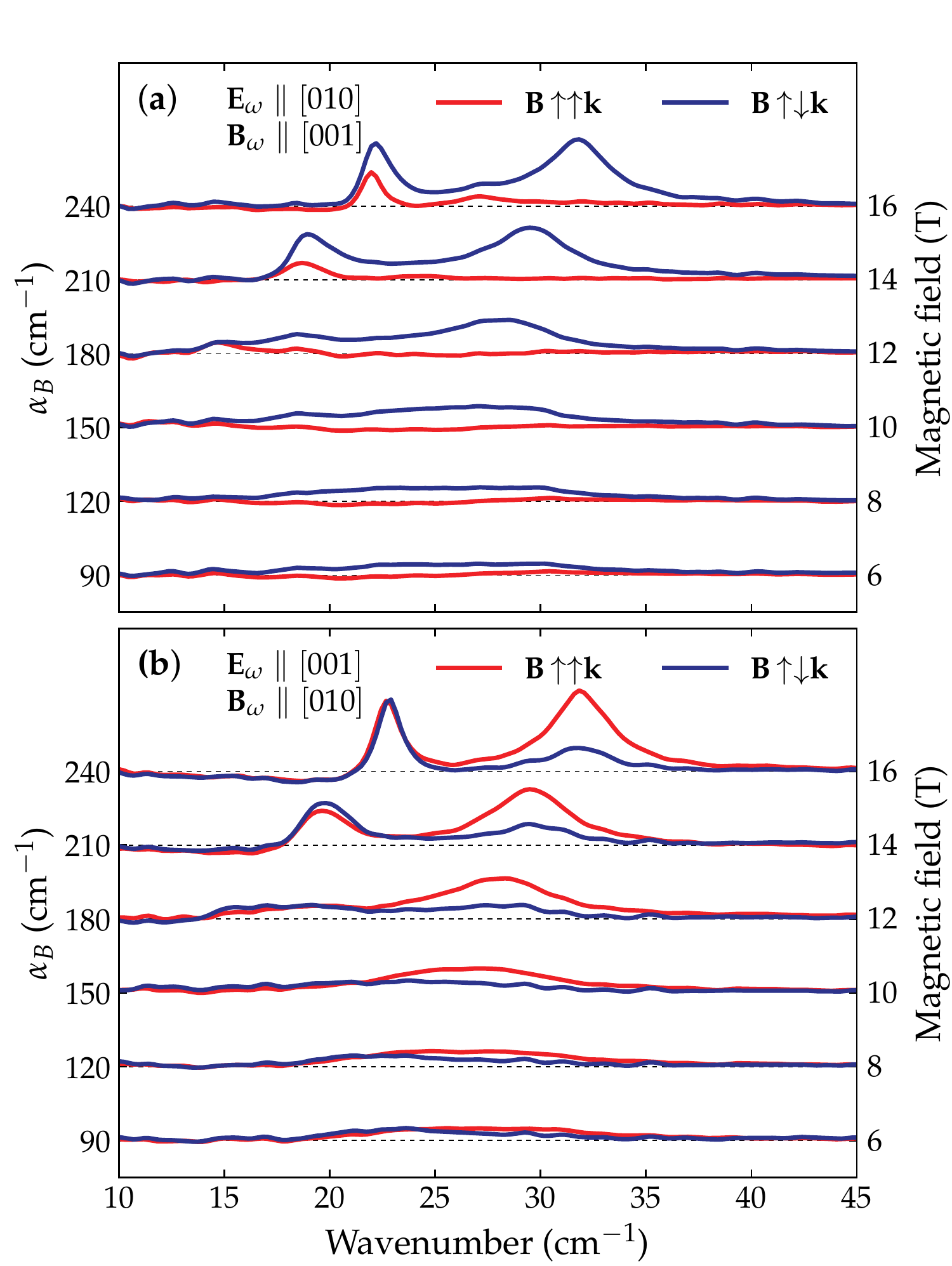}
	\caption{\label{fig:scso_b[100]_10k_experiment}
	Magnetic field dependence of absorption spectra of \scso/ in the paramagnetic state at 10\,K,  $ \bdc/\uparrow\uparrow \kvec/\parallel [100]$  (red) and $\bdc/\uparrow\downarrow \kvec/$  (blue). 
		The spectra are shifted in proportion to the absolute value of the magnetic field.
	}
\end{figure}

Fig.\,\ref{fig:scso_b[100]_14t_experiment}\,(a) and (b) show  the temperature dependence of  $\alpha^\pm_T$  between 3\,K and 100\,K in two polarizations of the THz radiation in magnetic field $\pm$14\,T. 
Below 7\,K, in the magnetically ordered phase,  the spectrum is dominated by three resonances at 18\wn/, 28\wn/ and 32\wn/. 
Since the resonance frequencies of the spin wave modes  are located at almost the same position\cite{Kezsmarki2014} in \scso/ and \bcgo/, and the magnetic field dependence of \Mvec/ and \Pvec/ are also similar in the two compounds,\cite{Murakawa2010,Akaki2012} we use the same assignment of spin waves as for  \bcgo/ .\cite{Penc2012} 
The 18\wn/ mode is the Goldstone mode of the easy-plane antiferromagnet gapped by the in-plane magnetic field whereas the latter two resonances correspond to the spin stretching modes. 
When the field is parallel or anti-parallel to the light propagation direction the spectra are markedly different. 
The absorption difference is  due to the MChD.\cite{Bordacs2012,Kezsmarki2014}

As the temperature increases the  spin-stretching modes at 28\wn/ and 32\wn/ merge  and eventually  disappear above 30\,K. 
However, the lowest energy mode,  the Goldstone mode in the ordered phase, is visible even at 100\,K. 
The MChD has pronounced polarization dependence in the paramagnetic phase. 
A strong MChD is observed  for all resonances in polarization $\eac/\parallel[010]$: the absorption coefficient is nearly zero for positive fields whereas finite absorption is detected for negative fields. 
In  the orthogonal polarization, $\eac/\parallel[001]$,  the lowest-energy resonance has  weak MChD and  changes  sign between  10 and 15\,K.

In 14\,T the MChD is the strongest at 10\,K,  close to the N\'eel temperature but  already in the paramagnetic phase, see Fig.\,\ref{fig:scso_b[100]_14t_experiment}\,(a) and (b).  
The magnetic field dependence of the absorption spectra  at 10\,K is shown in Fig.~\ref{fig:scso_b[100]_10k_experiment}\,(a) and(b).
The Goldstone-like mode suddenly appears above 12\,T and it only shows MChD for \eac/$\parallel$[010]. 
The  remnant spin stretching mode arises from the  broad absorption feature in low magnetic field and has a strong MChD with opposite signs in the two polarizations.

\begin{figure*}
	\includegraphics[width=0.95\textwidth]{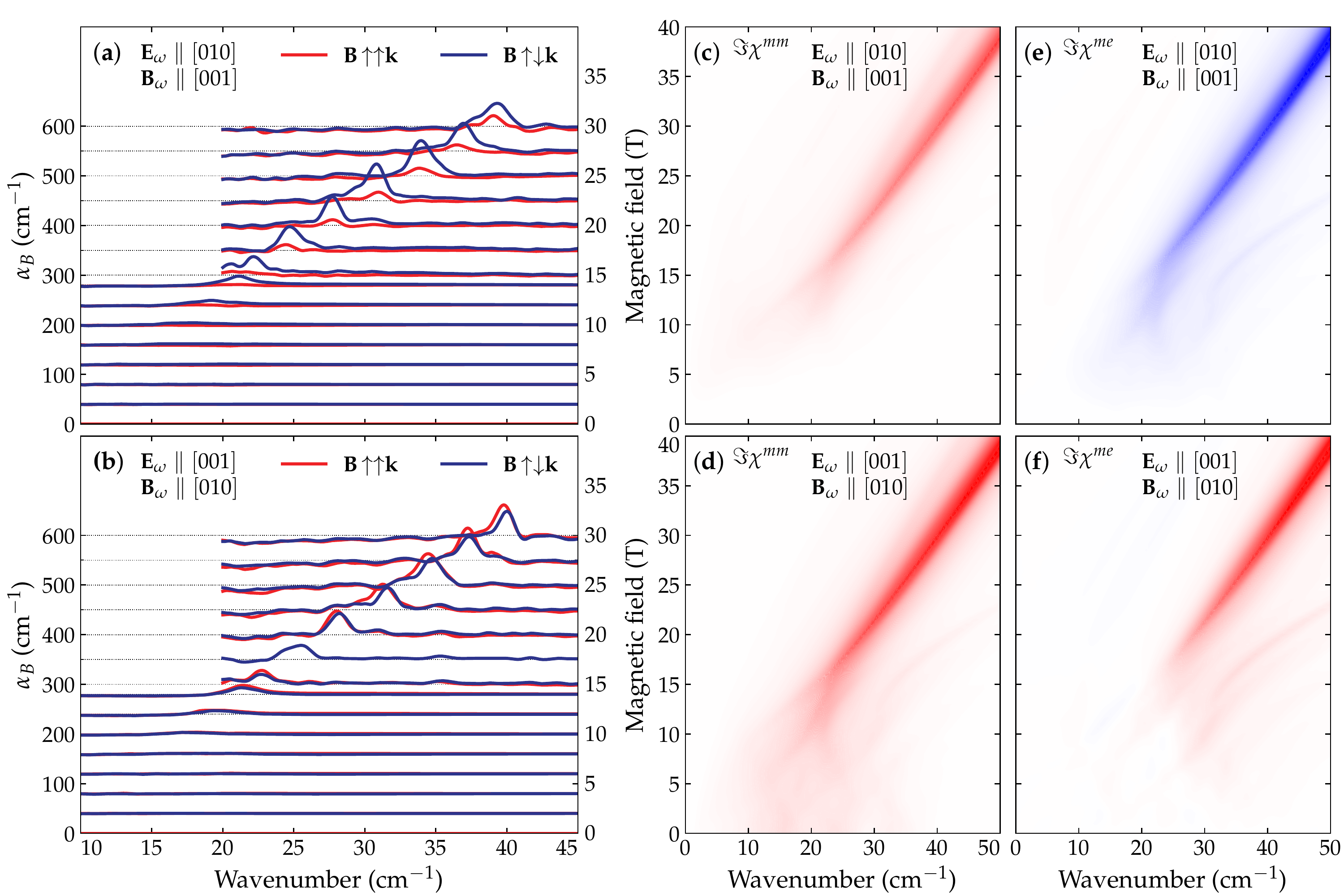}
	\caption{\label{fig:scso_b[100]_30k_experiment}
		Magnetic field dependence of measured THz absorption spectra (a-b) and calculated susceptibilities (c-f) of \scso/ at 30\,K.
		The THz absorption spectra for magnetic field $ \bdc/\uparrow\uparrow \mathbf{k}\parallel[100]$ (red) and $\bdc/\uparrow\downarrow \mathbf{k}$ (blue) are shown for two polarizations of the THz light. 
		The spectra are shifted in proportion to the absolute value of the magnetic field. 
		(c) and (e) are  the magnetic susceptibility $\Im\chi^{mm}(\omega)$  and the ME susceptibility $\Im\chi^{me}(\omega)$ for the polarization $\eac/\|[010]$ and $\bac/\|[001]$.
		(d) and (f) are   $\Im\chi^{mm}(\omega)$  and  $\Im\chi^{me}(\omega)$ for the polarization $\eac/\|[001]$ and $\bac/\|[010]$. 
		Red (positive) and blue (negative) colors indicate the sign of susceptibility. 
		The saturation of the color corresponds to the magnitude of the corresponding susceptibility matrix elements, $\Im \chi^{mm}(\omega)$, Eq.\,(\ref{eq:imchimm}), and $\Im \chi^{me}(\omega)$, Eq.\,(\ref{eq:imchime}). 
	The susceptibilities were calculated  by the exact diagonalization of a 4-site cluster.
	}
\end{figure*}

The magnetic field dependence of the absorption spectra at 30\,K is shown if Fig.\,\ref{fig:scso_b[100]_30k_experiment}\,(a) and (b).
The average intensity of the single resonance line, $(\alpha^+ + \alpha^-)/2$, observed at this temperature is nearly the same for both polarizations  and it grows  gradually as the field is increased. 
For polarization \eac/$\parallel$[010] a strong MChD is observed while only a small absorption difference is detected in \eac/$\parallel$[001].

\section{Spin Hamiltonian and magnetoelectric coupling}\label{sec:ED_spin_ham}

We consider the following spin Hamiltonian for $S=3/2$ Co$^{2+}$ spins coupled within a single layer, Fig.\,\ref{fig:2sublatt}, of \scso/:
\begin{multline}
	\mathcal{H}=J\sum_{(i,j)}\left( S^x_i  S^x_j+ S^y_i  S^y_j\right)+J_z \sum_{(i,j)} S^z_i  S^z_j+\Lambda\sum_{i}( S^z_i)^2\\
	+D_z\sum_{(i,j)}\left( S^x_i  S^y_j- S^y_i  S^x_j\right) - g \mu_\mathrm B {\bf B} \sum_{i}{\bf  S}_i,
	\label{eq:Hamiltonian}
\end{multline}
where $J$ and $J_z$ are the anisotropic exchange parameters, $\Lambda$ is the on-site anisotropy parameter, $D_z$ is the z component of the \DM/ vector and the last term is the Zeeman interaction.
 $(i,j)$ denotes pairs of nearest neighbor sites with $i\in \mathrm{A}$ and $j\in \mathrm{B}$, Fig.\,\ref{fig:2sublatt}. 
 The spin-nematic interactions responsible for the weak in-plane anisotropy are neglected.\cite{Soda2014,Romhanyi2011a}

From the symmetry point of view, the absence of inversion at the Co site allows the ME coupling. A suitable microscopic mechanism is provided by the spin-dependent $p$-$d$ hybridization. \cite{Arima2007} 
In the case of a tetrahedrally coordinated magnetic moment (the Co$^{2+}$ ion in our case),
the electric polarization is quadratic in spin components given by 
\cite{Murakawa2010, Miyahara2011}
\begin{eqnarray}
\mathbf{P} \propto \sum^4_{o=1}\left(\mathbf{S} \cdot\mathbf{e}_{o}\right)^2\mathbf{e}_{o},
\label{eq:poldef}
\end{eqnarray}
where  $\mathbf{e}_{o}$  is the unit vector pointing from the center of the (distorted) tetrahedron toward the four $o=1,\dots 4$ ligands (the oxygen ions) at the vertices of the tetrahedron. 

Since the Sr$_2$CoSi$_2$O$_7$ is composed of alternating tetrahedra, the polarization components are 
\begin{align}
 P^{x}_{j} &\propto
 -\cos2\kappa_j\left( S^x_{j} S^z_{j} \!+\! S^z_{j} S^x_{j}\right)
 -\sin2\kappa_j\left( S^y_{j} S^z_{j} \!+\! S^z_{j} S^y_{j}\right),
 \nonumber\\
 P^{y}_{j}&\propto
\cos2\kappa_j\left( S^y_{j} S^z_{j} \!+\! S^z_{j} S^y_{j}\right)
-\sin2\kappa_j\left( S^x_{j} S^z_{j} \!+\! S^z_{j} S^x_{j}\right),
\nonumber\\
 P^{z}_{j}&\propto
\cos2\kappa_j\left(( S^y_{j})^2 \!-\! ( S^x_{j})^2\right)
 -\sin2\kappa_j\left( S^x_{j} S^y_{j} \!+\! S^y_{j} S^x_{j}\right),
\label{eq:pol}
\end{align}
where $j$ belongs to either sublattice A, with a tilt angle $\kappa_j=\kappa$, or to sublattice B, where $\kappa_j=-\kappa$.\cite{Miyahara2011} 

The oscillating magnetic field of the light interacts with the total magnetization
\begin{equation}
\mathbf{M} =   \mathbf{M}_\mathrm A + \mathbf{M}_\mathrm B,
\label{eq:Mtot}
\end{equation}
while the electric field interacts with the total polarization 
\begin{equation}
\mathbf{P} = \mathbf{P}_\mathrm A + \mathbf{P}_\mathrm B,
\label{eq:Ptot}
\end{equation}
where the sublattice ($l = \mathrm A$ or $\mathrm B$) magnetization and polarization are
\begin{equation}
	\mathbf{M}_l = g \mu_\mathrm B  \sum_{j \in l}  \mathbf{S}_j  \quad\mathrm{and}\quad \mathbf{P}_l = \sum_{j \in l}  \mathbf{P}_j.
\end{equation}

We define the magnetic  susceptibility as
\begin{align}
	\chi^{mm}_{\mu\mu}(\omega) &=  \sum_{i,f} 
	\frac{\left|\langle f| M^\mu | i \rangle \right|^2}{\hbar\omega \!-\! E_i \!+\! E_f \!+\! i\delta} 
	\frac{e^{-\beta E_f} - e^{-\beta E_i}}{Z}, \label{eq:chimm}
\end{align} 
and the  ME susceptibility as
\begin{align}
	\chi^{me}_{\mu\nu}(\omega) &=  \sum_{i,f} \frac{
		\langle i | M^\mu | f \rangle \langle f | P^\nu | i \rangle}{\hbar\omega \!-\! E_i \!+\! E_f \!+\! i\delta}  
	\frac{e^{-\beta E_f} - e^{-\beta E_i}}{Z}, \label{eq:chime}
\end{align}
where $Z = \sum_i e^{-\beta E_i}$ is the partition sum, $\beta = 1/k_\mathrm B T$ is the inverse temperature, the $\delta$ parameter gives a finite broadening to the absorption peaks, and $E_i$ and $E_f$ are the energies of the initial and final spin states, respectively.

The experimentally measured NDD is related to the imaginary part of ME susceptibility\cite{Szaller2014} 
\footnote{The absorption is given by
	$$
	\alpha^\pm(\omega)= \frac{2\omega}{c_0}\Im \mathcal{N}_\pm(\omega),
	$$
	where the complex index of refraction $\mathcal{N}_\pm$ for the $\pm k$ direction of light propagation is
	$$
	\mathcal{N}_\pm(\omega) = \sqrt{\left[\epsilon_{\nu\nu}  + \chi^{\rm ee}_{\nu\nu}(\omega )\right] \left[\mu_{\mu\mu}+\chi^{\rm mm}_{\mu\mu}(\omega )\right]} \pm \chi^{me}_{\mu\nu}(\omega), $$
	where $\chi^{me}_{\mu\nu}(\omega)$ is the time-odd part of the ME susceptibility.  
	$\epsilon_{\nu\nu}$ and $\mu_{\mu\mu}$ are  real background dielectric and magnetic susceptibilities not originating from the spin system.
	In the limit of a small spin contribution to the susceptibilities, $\chi^{\rm ee}_{\nu\nu}(\omega)\ll \epsilon_{\nu\nu} $, $\chi^{\rm mm}_{\mu\mu}(\omega )\ll \mu_{\mu\mu}  $, index of refraction is 
\begin{eqnarray*}
	\mathcal{N}_\pm(\omega) &\approx& \sqrt{\epsilon_{\nu\nu}\mu_{\mu\mu}  } 
	+ \frac{\sqrt{\mu_{\mu\mu}  }}{2\sqrt{\epsilon_{\nu\nu} }}\,\chi^{\rm ee}_{\nu\nu}(\omega ) \\
	&&+
	\frac{\sqrt{\epsilon_{\nu\nu} }}{2\sqrt{\mu_{\mu\mu}}}\,\chi^{\rm mm}_{\mu\mu}(\omega )\pm \chi^{me}_{\mu\nu}(\omega).
\end{eqnarray*}
}
as
\begin{eqnarray}\label{eq:DeltaAlpha}
	\alpha^+(\omega) - \alpha^-(\omega) &=&\frac{4\omega}{c_0}\Im \chi^{me}_{\mu\nu}(\omega),
\end{eqnarray}
where $c_0$ is the speed of light in vacuum and frequency is in units of rad/s.

For a given transition $\ket{i}\rightarrow \ket{f}$ the imaginary (dissipative) 
parts of the magnetic and magnetoelectric susceptibilities are:\cite{Szaller2014}
\begin{eqnarray}
	\Im \chi^{mm}_{\mu \mu} (\omega)  &\propto & \left| \bra{i} M^\mu \ket{f} \right|^2  \delta(\omega-\omega_{if}),\label{eq:imchimm}\\
	\Im \chi^{me}_{\mu \nu} (\omega)  &\propto &\Re  \left\{ \bra{i} M^\mu \ket{f}  \bra{f} P^\nu \ket{i} \right\} \delta(\omega-\omega_{if}),
	\label{eq:imchime}
\end{eqnarray}
where $\omega_{if}=(E_f-E_i)/\hbar $ is the transition frequency and  $\omega$  is the frequency of a photon.

\section{Exact diagonalization}\label{sec:ED}

To get a first insight into the nature of excitations, we performed an exact diagonalization study of a small cluster containing four Co$^{2+}$ ions, i.e. two unit cells, at finite temperatures.
\footnote{
	Since the $S=3/2$ has 4 states, the dimension of the Hilbert space  of the 4-spin cluster  is $4^4 = 256$, and the sums in Eqs.~(\ref{eq:chimm}) and (\ref{eq:chime}) run over $256^2 = 65536$ terms. 
	Even though the cluster is small, it already provided sufficient information to compare with the experimental data at high temperatures and fields.} 
We note that exact diagonalization was also used to study the ME excitations of \bcgo/ at zero temperature in Ref.~[\onlinecite{Miyahara2011}].

Since the magnetization curves closely follow those of \bcgo/,\cite{Murakawa2010,Akaki2012} we assume  the same set of parameters describes both,  \bcgo/ and \scso/. 
We use the parameters obtained from absorption spectra of \bcgo/ in Ref.~[\onlinecite{Penc2012}],
 {\it i.e.} $\Lambda = 13.4$\,K, $J= 2.3$\,K, $J_{zz} = 1.8$\,K, $D_z=-0.1$, $\kappa=22.4^\circ$, and $g=2.3$. 
This set of parameters provided a remarkable good agreement with the experimentally measured absorptions. 
The results of these calculations are presented in Fig.\,\ref{fig:scso_b[100]_14t_experiment}(c)-(f) and Fig.\,\ref{fig:scso_b[100]_30k_experiment}(c)-(f). 

The exact diagonalization results  show,  in accordance with the experiment, that a single absorption line is present in  high magnetic fields  at high temperatures and a second resonance appears as the temperature is lowered below 20\,K. 
Calculation predicts a finite ME effect responsible for the observed NDD. 
Moreover, the sign change of NDD observed for  $\eac/\parallel[001]$ between 10 and 15\,K is reproduced by the numerical calculations, Fig.~\ref{fig:scso_b[100]_14t_experiment}(b) and (f).

\section{The single-ion problem}\label{sec:single_ion}

In order to get a deeper understanding of the THz absorption spectra and NDD in the paramagnetic phase of  \scso/, we consider a model of a single Co$^{2+}$ spin in the center of a tetrahedron. 
This unit is the building block of the crystal lattice of \scso/, and at the same time, its symmetry is representative of the field-induced reduction of the point symmetry in the real material. 
More precisely, in the next two sections we will show that the magnetic space group at the $\Gamma$-point is isomorphic to the magnetic point group of the tetrahedron. 
Furthermore,  such a simple single-ion model describes the essence of the experiments in the disordered paramagnetic state where the strong thermal fluctuations smear out the coupling to  the neighboring spins.

\subsection{The Hamiltonian and the symmetries}
\subsubsection{The spin Hamiltonian and the electric polarization}

\begin{figure}[tb]
\includegraphics[width=0.8\columnwidth]{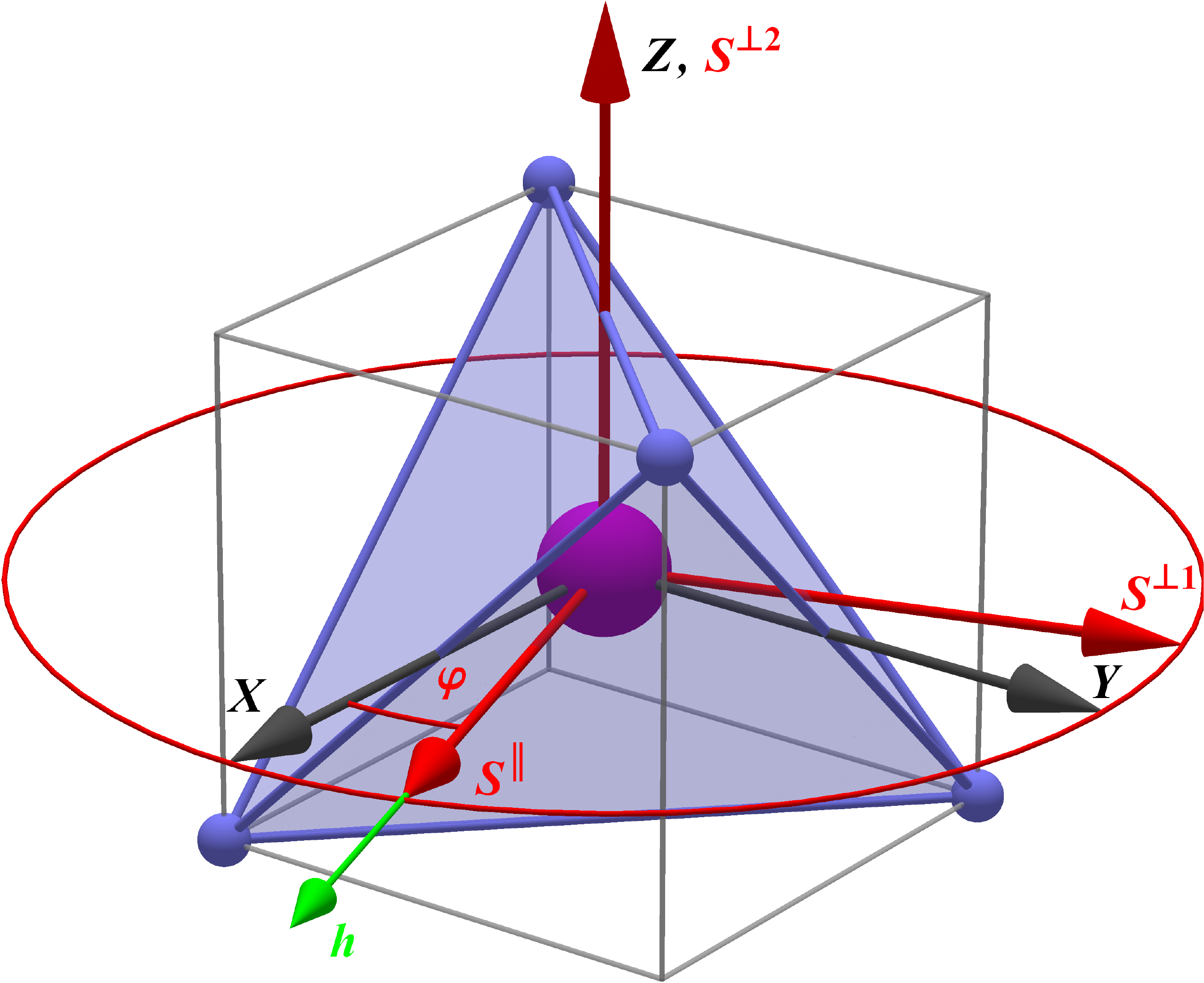}
\caption{The tetrahedron-fixed coordinate system $\{{X,Y,Z}\}$ and the rotated  magnetic-field-fixed coordinate system $\{\parallel, \perp\!1,\perp\!2\}$. 
Purple ball is the   Co$^{2+}$ and the blue tetrahedron is the  cage with the oxygen ions at the vertices. 
The  quantization  axis (spin component $S^{\parallel}$) is parallel to  the direction of the  static external magnetic field $h$ (shown in green), while  one perpendicular axis (the spin component $S^{\perp 2}$) is parallel to  $Z$.
\label{fig:3Dcoords}
}
\end{figure}

We will  consider  magnetic field in the $XY$-plane of a single tetrahedron with a coordinate system defined in Fig.\,\ref{fig:3Dcoords}.
It is convenient to use the coordinate system where the ${\parallel}$ axis points in the direction of the field  $h = g \mu_\mathrm B B$, the ${\perp\!2}$ axis is parallel to the $Z$ direction and the ${\perp\!1}$ direction is chosen so that the three axes form an orthogonal right-handed system.
The spin components in the coordinate system of the tetrahedron are related to the spin components in the field-fixed system as
\begin{eqnarray}
S^{X}&=&\cos \varphi \; S^{\parallel} - \sin \varphi \; S^{\perp 1},\nonumber\\
S^{Y}&=&\sin \varphi \; S^{\parallel} + \cos \varphi \; S^{\perp 1},\\
S^{Z}&=&S^{\perp 2}.\nonumber
\end{eqnarray}

Following Eq.~(\ref{eq:Hamiltonian}) the  Hamiltonian
for a single spin in the field-fixed coordinates is
\begin{equation}
\mathcal{H}= \Lambda\left(S^{\perp 2}\right)^2 - h  S^{\parallel} \;.
\label{eq:H_1ion}
\end{equation}
The quantization axis  is chosen along the magnetic field with eigenvalues and eigenvectors of $S^\parallel$ being
$\{+3/2,+1/2,-1/2,-3/2\}$ and  $\{\Uparrow,\uparrow,\downarrow,\Downarrow\}$, respectively. 

The magnetic field and the anisotropy lower the $O(3)$ spatial symmetry to the group generated by inversion and a twofold rotation $C_2^{\parallel}$ about the magnetic field. 
Time reversal symmetry alone is broken, but time reversal followed by  a rotation perpendicular to the field -- denoted by $\Theta C_2^{\perp 2}$ -- remains a symmetry element.

From  Eq.~(\ref{eq:poldef}) we get for the electric polarization
\begin{eqnarray}
P^{X}&=&\eta_{XY} (S^Z S^Y + S^Y S^Z),\nonumber\\
P^{Y}&=&\eta_{XY} (S^Z S^X + S^X S^Z), \label{eq:pop1}\\
P^{Z}&=&\eta_{Z}(S^X S^Y + S^Y S^X),\nonumber
\end{eqnarray}
where for the regular tetrahedron $\eta_{XY}=\eta_Z$. 
 The relation  to the components in the field-fixed coordinates is 
\begin{eqnarray}
P^{X}&=&\cos \varphi \; P^{\parallel} - \sin \varphi \; P^{\perp 1},\nonumber\\
P^{Y}&=&\sin \varphi \; P^{\parallel} + \cos \varphi \; P^{\perp 1},\\
P^{Z}&=&P^{\perp 2}.\nonumber
\end{eqnarray}

 \subsubsection{Solution of the Hamiltonian}

\begin{figure}[tb]
\includegraphics[width=0.45\textwidth]{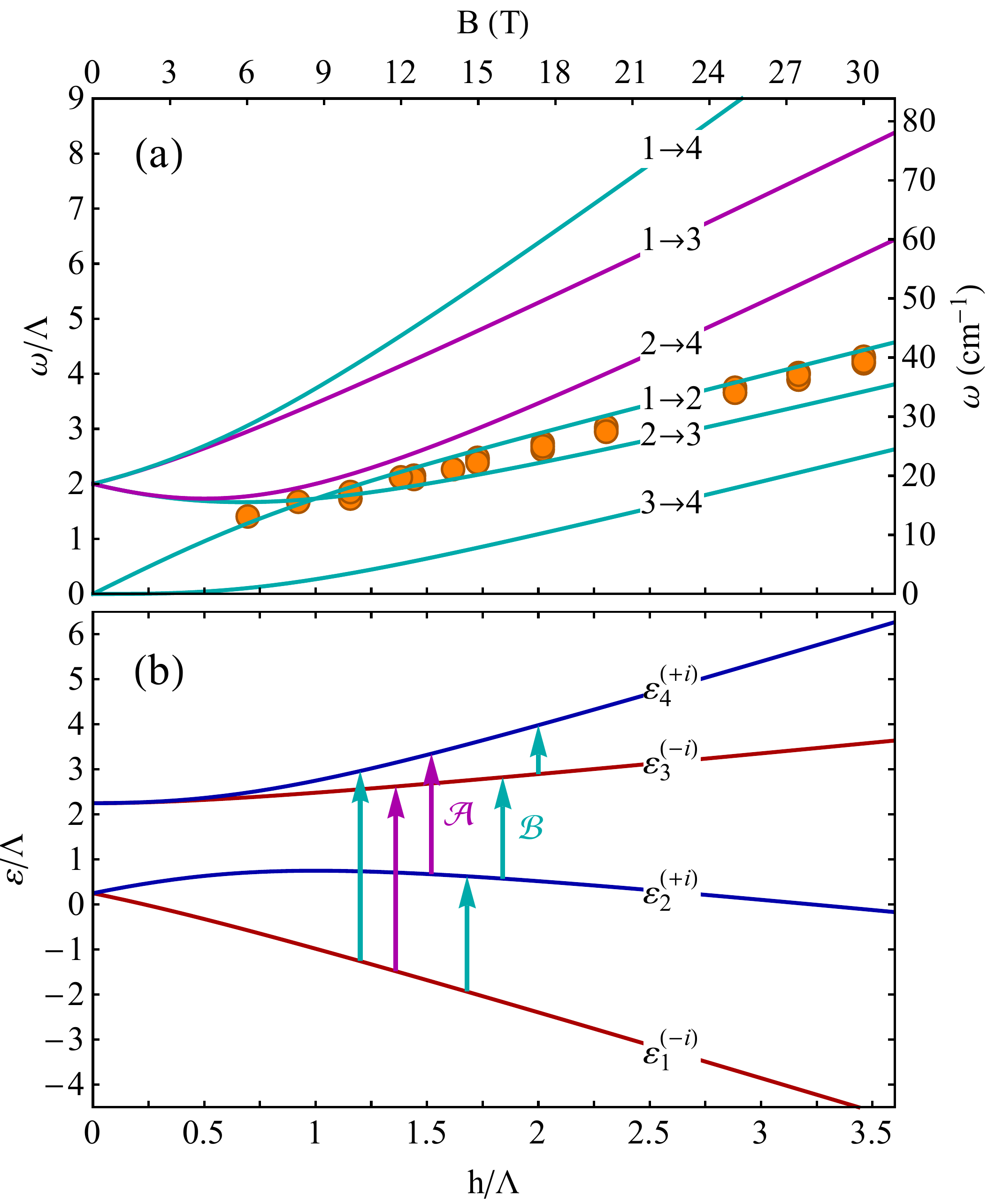}
\caption{\label{fig:single_ion_e_de}
(a) Transition energies $\omega$ and (b) energy levels $\varepsilon$ of the single-ion model with easy plane anisotropy $\Lambda$ in a magnetic field $h = g \mu_\mathrm B B$ within the easy plane as a function of $h/\Lambda$. States 1 and 3 (red curves) are multiplied by $-i$ after a $\pi$-rotation about the field ($C_2^{\parallel}$), while states 2 and 4 (blue) get an $i$. 
Magenta arrows represent transitions between the states of the same symmetry, induced by operator $\mathcal{A}$, even under the $C_2^{\parallel}$, such as $S^{\parallel}$, $P^{\parallel}$, see Table \ref{table:chartab_double_C2z}.
Cyan arrows connect states with different symmetries induced by operator $\mathcal{B}$, odd under $C_2^{\parallel}$, such as  the perpendicular components of  spin and polarization operator.
The color of the curves corresponds to the color of the arrows in panel (b). 
The filled circles show the  experimental results.
}
\end{figure}

The matrix representation of the Hamiltonian (\ref{eq:H_1ion}) in the spin  basis $\{\Uparrow,\uparrow,\downarrow,\Downarrow\}$ is 
\begin{equation}
\mathcal{\hat{H}}=
\left(
\begin{array}{cccc}
 \frac{3}{4} \Lambda \!-\! \frac{3}{2} h   & 0 & -\frac{\sqrt{3}}{2} \Lambda & 0 \\
 0 & \frac{7}{4} \Lambda \!-\! \frac{1}{2} h & 0 & -\frac{\sqrt{3}}{2} \Lambda  \\
 -\frac{\sqrt{3}}{2} \Lambda  & 0 & \frac{7}{4} \Lambda \!+\! \frac{1}{2} h & 0 \\
 0 & -\frac{\sqrt{3}}{2} \Lambda  & 0 & \frac{3}{4} \Lambda \!+\! \frac{3}{2} h  \\
\end{array}
\right).\\
\end{equation}

The eigenvalues $\pm i$ of the rotation operator $\hat{C}_2^{\parallel}$, 
\begin{equation}
\hat{C}_2^{\parallel}=e^{i \pi \hat{S}^\parallel}=
\left(
\begin{array}{cccc}
-i & 0 & 0 & 0 \\
0 & i & 0 & 0 \\
0 & 0 & -i & 0 \\
0 & 0 & 0 & i \\
\end{array}
\right)
\label{eq:C2parmat}
\end{equation}
are good quantum numbers as the operator commutes with the Hamiltonian, $[\hat{\mathcal{H}},\hat{C}_2^{\parallel}]=0 $.
Therefore, only the states with the same eigenvalue of $\hat{C}_2^{\parallel}$ are mixed.
We use $\pm i$ to label the eigenstates, $|\psi_j^{\left(\pm i\right)} \rangle$ and corresponding energies, $\varepsilon_j^{(\pm i)}$.
 
The  energies in  increasing order  are
\begin{subequations}
\label{eq:1ion_energies}
\begin{align}
\varepsilon_1^{(-i)} &= -\frac{h}{2}+\frac{5 \Lambda }{4}-\sqrt{h^2+ h \Lambda +\Lambda ^2}, \\
\varepsilon_2^{(+i)} &=\frac{h}{2}+\frac{5 \Lambda }{4}-\sqrt{h^2- h \Lambda +\Lambda ^2},\\
\varepsilon_3^{(-i)} &=-\frac{h}{2}+\frac{5 \Lambda }{4}+\sqrt{h^2+ h \Lambda +\Lambda ^2},\\
\varepsilon_4^{(+i)} &=\frac{h}{2}+\frac{5 \Lambda }{4}+\sqrt{h^2-g h \Lambda +\Lambda ^2},
\end{align}
\end{subequations}
and the corresponding unnormalized eigenstates are
\begin{subequations}
\label{eq:eigen}
\begin{align}
|\psi_1^{(-i)} \rangle &\propto \left(2 h \!+\! \Lambda \!+\! 2 \sqrt{h^2 \!+\!  h \Lambda  \!+\! \Lambda ^2}\right) \left|\Uparrow\right\rangle + \sqrt{3} \Lambda \left|\downarrow\right\rangle , \label{eq:eigen1}\\
|\psi_2^{(+i)} \rangle &\propto \left(2 h \!-\! \Lambda \!+\! 2 \sqrt{h^2 \!-\!  h \Lambda  \!+\! \Lambda ^2}\right) \left|\uparrow\right\rangle + \sqrt{3} \Lambda \left|\Downarrow\right\rangle , \label{eq:eigen2}\\
|\psi_3^{(-i)} \rangle &\propto \left(2 h \!+\! \Lambda \!+\! 2 \sqrt{h^2 \!+\!  h \Lambda  \!+\! \Lambda ^2}\right) \left|\downarrow\right\rangle - \sqrt{3} \Lambda \left|\Uparrow\right\rangle , \label{eq:eigen3}\\
|\psi_4^{(+i)} \rangle &\propto \left(2 h \!-\! \Lambda \!+\! 2 \sqrt{h^2 \!-\!  h \Lambda  \!+\! \Lambda ^2}\right) \left|\Downarrow\right\rangle - \sqrt{3} \Lambda \left|\uparrow\right\rangle .\label{eq:eigen4}
\end{align}
\end{subequations}
The phases for the eigenvectors above are chosen in such a way that we recover the basis $\left\{\left|\Uparrow\right\rangle, \left|\uparrow\right\rangle, \left|\downarrow\right\rangle, \left|\Downarrow\right\rangle \right\}$ for $h \gg \Lambda $, e.g. $|\psi_1^{(-i)} \rangle \to \left|\Uparrow\right\rangle$, and so on.

\subsubsection{Transition matrix elements} 
\begin{table}[tb]
	\caption{\label{table:chartab_double_C2z} Character table of the double group corresponding to the group $C_{2}$, including the transformation properties of the operators and states. The group element $\mathbf{\bar 1}$  is the $2\pi$ rotation with the property $\mathbf{\bar 1}^2=\mathbf{1}$.}
	\begin{ruledtabular}
		\begin{tabular}{ c c c c c c}
			Irrep&$\mathbf{1}$&$C_2^{\parallel}$&$ \mathbf{\bar 1} $&${\bar C}_2^{\parallel} $&Operators\\ \hline
			$\mathsf{A}$&$1$&$1$&$1$&$1$&$S^{\parallel}$, $P^{\parallel}$\\
			$\mathsf{B}$&$1$&$-1$&$1$&$-1$&$S^{\perp 1}$, $S^{\perp 2}$, $P^{\perp 1}$, $P^{\perp 2}$\\
			$\mathsf{\overline{E}_1}$&$1$&$i$&$-1$&$-i$& $\left|\Downarrow\right\rangle$,  $\left|\uparrow\right\rangle$, $\left\langle \Uparrow\right|$,  $\left\langle \downarrow\right|$\\
			$\mathsf{\overline{E}_2}$&$1$&$-i$&$-1$&$i$& $\left|\Uparrow\right\rangle$,  $\left|\downarrow\right\rangle$,  $\left\langle \Downarrow\right|$,  $\left\langle \uparrow\right|$\\
		\end{tabular}
	\end{ruledtabular}
\end{table}

Based on the transformation properties under the rotation $C_2^\parallel$, as summarized in Table \ref{table:chartab_double_C2z}, we can construct selection rules for the matrix elements of the spin and polarization components. 
A matrix element for an operator $\mathcal{O}$ between states $\ket{\psi_\alpha}$ and $\ket{\psi_\beta}$
can only be nonvanishing if it transforms according to the totally symmetric $\mathsf{A}$ irrep of the group, $\mathsf{A}\subseteq\Gamma^{\beta}\otimes \Gamma(\mathcal{O}) \otimes \Gamma^{\alpha} $. 
In order to describe the transformation properties of the states of a half-integer spin  we need to consider the double group corresponding to the group $C_2$.  
If an operator $\mathcal{A}$ transforms as $\mathsf{A}$ and an operator  $\mathcal{B}$ as $\mathsf{B}$, and the spin states as $\mathsf{\overline{E}_1}$ and $\mathsf{\overline{E}_2}$ of the double group of $C_2$, Table~\ref{table:chartab_double_C2z},  the nonvanishing matrix elements are 
\begin{equation}
\label{eq:AB_selection_rule}
\left\langle \psi_\alpha^{(\pm i)}\right|\mathcal{A}\left|\psi_\beta^{(\pm i)}\right\rangle \quad\textrm{and}\quad
\left\langle \psi_\alpha^{(\mp i)}\right|\mathcal{B}\left|\psi_\beta^{(\pm i)}\right\rangle.
\end{equation} 
Allowed transitions between states of the same symmetry are of type $\mathcal{A}$ and allowed transitions between states of different symmetry are of type $\mathcal{B}$. We show these  transitions of the single-ion model in Fig.~\ref{fig:single_ion_e_de}.

In what follows, we take into account the effects of antiunitary symmetries containing the time-reversal operation $\Theta$ on the matrix elements. 
We show in the Appendix\,\ref*{sec_TBD} that any linear operator $\hat{\mathcal{O}}$, if the symmetry operation $\Theta C_2^{\perp 2}$ is present (as in the case of a magnetic field applied in the $XY$ plane), must satisfy $\hat{\mathcal{O}}=\pm \hat{\mathcal{O}}^*$.
Thus, the matrix elements of any linear operator even (odd) under the symmetry operation $\Theta C_2^{\perp 2}$ are real (imaginary). 

In conclusion, the unitary symmetries of the single-ion model determine the selection rules, i.e.~the non-vanishing matrix elements, whereas the antiunitary symmetry  constrains the matrix element to a real or to an imaginary value.

\subsection{Directional dichroism in the single-ion model}

Imaginary part of the ME susceptibility $\Im  \chi^{me}_{\mu \nu} (\omega)  $ causes NDD,\cite{Barron1984,Kezsmarki2011,Bordacs2012} Eq.\,(\ref{eq:DeltaAlpha}). 
For a given transition $\ket{i}\rightarrow \ket{f}$ the dissipative part of the ME susceptibility is proportional to the real part of the matrix element product, $\Re \{\bra{f}M^\mu\ket{i}\bra{f}P^\nu\ket{i}\}$, Eq.\,\ref{eq:imchime}.
Thus, NDD is non-zero if both matrix elements are non-zero for the same pair of states  $\ket{i}$ and $\ket{f}$.
This is allowed by symmetry  if $M^\mu$ and $P^\nu$ transform according to the same irreducible representation of the group of unitary symmetries.

Furthermore, the product of the $M^\mu$ and $P^\nu$ matrix elements has a finite real part only if both matrix elements are real or both are imaginary.
Thus, both operators must be even ($\mathcal{A}$ type) or both odd ($\mathcal{B}$ type) under antiunitary symmetry operation, as was shown in the previous section.
The summary of all spin and polarization operator amplitudes for  $S=3/2$ in the single-ion model is presented in Table~\ref{table:selection rules}. 
They are either real, imaginary, or symmetry-forbidden.

Below we  examine the symmetry properties of the operators for  different directions of the external field with respect to the  cobalt-oxygen tetrahedron in the easy plane to find out the details of the existence of NDD.

\begin{figure}[tb]
\includegraphics[width=0.9\columnwidth]{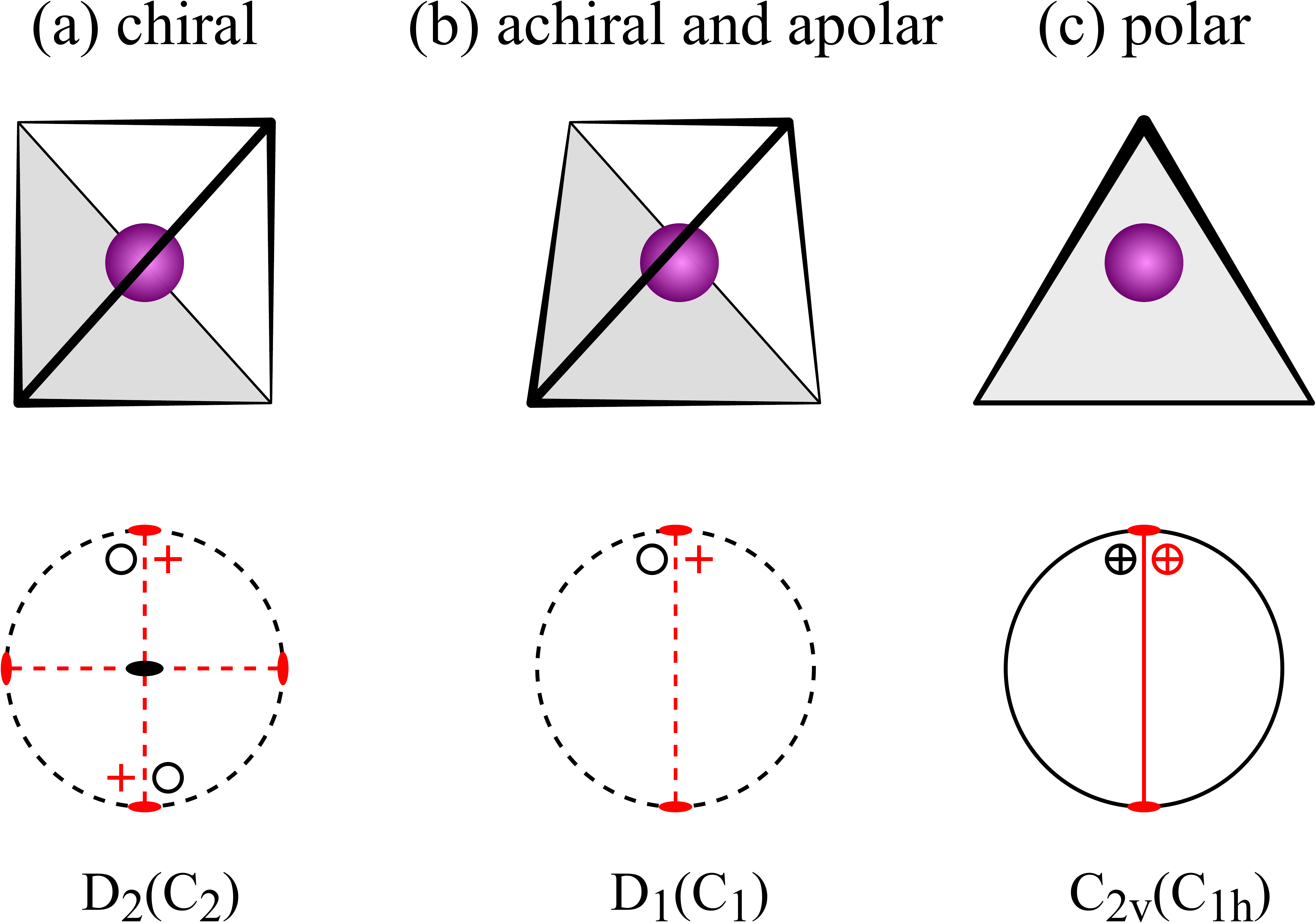}
\caption{ The tetrahedron in the (a) chiral ($\varphi=0$), (b) low symmetry, and (c) polar ($\varphi=\pi/4$) cases as seen from the direction of the magnetic field $\mathbf{B}$. The magenta sphere represents the magnetic ion at the center of tetrahedral cage. The stereograms represent the magnetic point group for each case, black refers to the unitary subgroup and red the  antiunitary part combined with the time reversal.  \label{fig:two_mapping}
}
\end{figure}

\subsubsection{Magneto-chiral dichroism (chiral case): $\varphi=0$}
\label{sec:chiral}

When the external field is parallel to one of the two-fold rotation axes, $\varphi=0$ (see Fig.~\ref{fig:3Dcoords}), the system has  chiral symmetry $D_{2}\left(C_{2}\right)$ in Schoenflies notation (see Fig.~\ref{fig:two_mapping}(a)), and the polarization operators in the local coordinate system are
\begin{subequations}
\begin{align}
P^{\parallel}_\text{chiral} &= \frac{\eta_{XY}}{2 i}\left[(S^+)^2-(S^-)^2 \right],
\label{eq:P_par_chiral}
\\
P^{\perp 1}_\text{chiral} &= \frac{\eta_{XY}}{2 i}\left[ S^\parallel( S^+ -S^-) + (S^+ -S^-)S^\parallel \right],
\\
P^{\perp 2}_\text{chiral} &=  \frac{\eta_Z}{2}\left[ S^\parallel( S^+ +S^-) + (S^+ +S^-)S^\parallel \right],
\end{align}
\end{subequations}
where $S^\pm = S^{\perp 1} \pm i S^{\perp 2}$.
This is also the case (up to a sign) when $\varphi$ is an integer multiple of $\pi/2$ due to the $S_4$ symmetry of the distorted oxygen tetrahedron. 
It is worth noting, that the perpendicular components  $P^{\perp 1}$ and $P^{\perp 2}$ change the $S^\parallel$ quantum number  by $\pm$1, creating dipolar spin excitation, but $P^{\parallel}_\text{chiral}$ changes  by $\pm$2,  creating quadrupolar spin excitation.\cite{Akaki2017}

None of the components of \Pvec/ transforms according to the fully symmetric irreducible representation $\mathsf{A_+}$,   Table\,\ref{table:chartab_chiral_polar}. Therefore  the expectation value for the static  polarization is zero in the ground state.
Although the chiral case is apolar,   the dynamic ME susceptibility is allowed.
The operators $S^{\perp 2}$ and $P^{\perp 1}$ belong to the same irreducible representation $\mathsf{B_-}$ of $D_{2}\left(C_{2}\right)$, thus the dynamic ME susceptibility is allowed. 
As the oscillating magnetization $S^{\perp 2}$ and polarization $P^{\perp 1}$  are perpendicular to each other and to the external magnetic field, we expect NDD in the Faraday geometry, when the light is propagating parallel to the field. 
This effect is nothing else but the MChD. 
The operators $S^{\perp 1}$ and $P^{\perp 2}$, corresponding to the perpendicular polarization of radiation as compared to $S^{\perp 2}$ and $P^{\perp 1}$,  belong to the same  irreducible representation $\mathsf{B_+}$ and therefore the  NDD will appear irrespective of polarizations of the incident light in Faraday geometry. 
The symmetry argument presented above is supported by the direct evaluation of the matrix elements (see Appendix \ref{sec_matrix_elements} and Table\,\ref{table:selection rules}) using the eigenstates given in Eqs.\,(\ref{eq:eigen1})-(\ref{eq:eigen4}). 

\begin{table*}
	\caption{\label{table:selection rules} Matrix elements of the spin and polarization operators, first in small field $h$, then for small anisotropy $\Lambda$, in the single-ion model. 
		The first non-vanishing order in either $h$ or  $\Lambda$ is shown. 
		$ c$ stands for a real constant.
	}
	\begin{ruledtabular}
		\begin{tabular}{ c c c c c c c c}
			transition&$S^{\parallel}$&$S^{\perp 1}$&$S^{\perp 2}$ & $P^{\parallel}_\text{chiral},P^{\perp 1}_\text{polar}$&$P^{\perp 1}_\text{chiral},P^{\parallel}_\text{polar}$&$P^{\perp 2}_\text{chiral}$&$P^{\perp 2}_\text{polar}$\\ 
			\hline
			$1\to 2$& $-$         &$c$, $c$        &$ic$, $ic$        &$-$    &$ih$, $ic$      &$h$, $c$       &$-$\\
			$2\to 3$& $-$         &$c$, $c$       &$ih$, $ic$        &$-$    &$ic$, $i\Lambda$&$c$, $\Lambda$ &$-$\\
			$3\to 4$& $-$         &$h^2$, $c$     &$ic$, $ic$     &$-$    &$ih$, $ic $     &$h$, $c$       &$-$\\
			$1\to 3$&$c$, $\Lambda$& $-$             &$-$          &$ic$, $ic$&$-$        &$-$           &$c$, $c$ \\
			$2\to 4$&$c$, $\Lambda$& $-$             &$-$          &$ic$, $ic$&$-$        &$-$           &$c$, $c$ \\
			$1\to 4$& $-$         &$c$, $\Lambda^2$&$ih$, $i\Lambda^2$&$-$    &$ic$, $i\Lambda$&$c$, $\Lambda$ &$-$\\
		\end{tabular}
	\end{ruledtabular}
\end{table*}

\begin{table}[tb]
\caption{\label{table:chartab_chiral_polar} Character table for the  magnetic point groups and symmetry-allowed operators for three orientations of the applied magnetic field in the easy plane, see Fig.\,\ref{fig:3Dcoords},  creating chiral ($\varphi=0$), polar ($\varphi=\pi/4$) and low symmetry cases.
}
\begin{ruledtabular}
\begin{tabular}{ c c c c c c c}
	\multicolumn{7}{c}{ $\varphi=0$: chiral case, the group is  $D_{2}(C_{2})$}\\
Irrep.&$\mathbf{1}$&$C_2^{\parallel}$&$\Theta C_2^{\perp 2}$&$\Theta C_2^{\perp 1} $&Operator(s)&NDD\\ \hline
$\mathsf{A_+}$&$ 1$&$ 1$&$ 1$&$ 1$&$S^{\parallel}$&--\\
$\mathsf{A_-}$&$ 1$&$ 1$&$-1$&$-1$&$P^{\parallel}$&--\\
$\mathsf{B_+}$&$ 1$&$-1$&$ 1$&$-1$&$S^{\perp 1}$, $P^{\perp 2}$&Faraday\\
$\mathsf{B_-}$&$ 1$&$-1$&$-1$&$ 1$&$S^{\perp 2}$, $P^{\perp 1}$&Faraday  \\ 
\hline 
\vspace{-6pt}
\\
\multicolumn{7}{c}{$\varphi=\pi/4$ : polar case, the group is $C_{2v}(C_{1h})$}\\
&$\mathbf{1}$&$\sigma^{\parallel}$&$\Theta C_2^{\perp 2}$&$\Theta \sigma^{\perp 1} $&\\ \hline
$\mathsf{A_+}$&$1$&$ 1$&$ 1$&$ 1$&$S^{\parallel}$, $P^{\perp 2}$&Voigt\\
$\mathsf{A_-}$&$1$&$ 1$&$-1$&$-1$&$P^{\perp 1}$&--\\
$\mathsf{B_+}$&$1$&$-1$&$ 1$&$-1$&$S^{\perp 1}$&--\\
$\mathsf{B_-}$&$1$&$-1$&$-1$&$ 1$&$S^{\perp 2}$, $P^{\parallel}$&Voigt \\ %
\hline 
\\
\multicolumn{7}{c}{$\varphi\neq 0$  nor $\pi/4$ :  the group is $D_1(C_1)$}\\
&$\mathbf{1}$&&$\Theta C_2^{\perp 2}$&  \\ 
\hline
$\mathsf{\Gamma_+}$&$1$&&$ 1$&&$S^{\parallel}$, $S^{\perp 1}$, $P^{\perp 2}$&both\\
$\mathsf{\Gamma_-}$&$1$&&$-1$&&$S^{\perp 2}$, $P^{\parallel}$, $P^{\perp 1}$&both\\
\end{tabular}
\end{ruledtabular}
\end{table}

From Table~\ref{table:chartab_chiral_polar} we can also infer the selection rules with regard to the quantum numbers $\pm i$, the eigenvalues of the  rotation operator $C_2^{\parallel}$. 
The $S^{\parallel}$ and $P^{\parallel}$ are invariant under $C_2^{\parallel}$, thus, they excite only type $\mathcal{A}$ transitions between states that have the same quantum number (see Fig.\,\ref{fig:single_ion_e_de} (b)), i.e. between states 1 and 3  and between states 2 and 4.
However, there is no  NDD for the $\mathcal{A}$ type transitions as $P^\parallel$ and $S^\parallel$ do not transform according to the same irrep.

The perpendicular components of $\mathbf{P}$ and $\mathbf{S}$ have matrix elements between states with different $C_2^{\parallel}$ quantum numbers, which corresponds to type $\mathcal{B}$ transitions, i.e. transitions $1\to 2$,  $1\to 4$, $2\to 3$ , and $3\to 4$ (see Fig.~\ref{fig:single_ion_e_de} (b)). 
We also see, Table\,\ref{table:chartab_chiral_polar},  that $S^{\perp 2}$ and  $P^{\perp 1}$ are both odd under $\Theta C_2^{\perp 2}$, with imaginary matrix elements, and their product is real in Eq.~(\ref{eq:imchime}), providing a finite imaginary $\chi^{me}_{\perp 2,\perp 1} (\omega)$ and a finite MChD. 
Similarly, the $S^{\perp 1}$ and $P^{\perp 2}$ are both even, with real matrix elements, providing a finite imaginary $\chi^{me}_{\perp 1,\perp 2} (\omega)$. 
In any other configuration the $\Im \left\{ \chi^{me}_{\mu \nu} (\omega)  \right\}=0$. 
Therefore, the NDD is present only in the Faraday configuration, $\kvec/\parallel \bdc/$. 

\subsubsection{Toroidal dichroism (polar case): $\varphi=\frac{\pi}{4}$ }
\label{sec:polar}
If the field direction is parallel to the upper edge of the tetrahedron, $\varphi=\frac{\pi}{4}$ in Fig.~\ref{fig:3Dcoords}, and the polarization operators are
\begin{subequations}
\begin{align}
P^{\parallel}_\text{polar} &= \frac{\eta_{XY}}{2 i}\left[ S^\parallel( S^+ -S^-) + (S^+ -S^-)S^\parallel \right], \label{eq:P_par_polar}\\
P^{\perp 1}_\text{polar} &= \frac{\eta_{XY}}{2 i}\left[(S^-)^2-(S^+)^2 \right],
\label{eq:P_perp1_polar}
\\
P^{\perp 2}_\text{polar} &= \frac{\eta_Z}{4}\left[4 (S^\parallel)^2  - (S^-)^2 - (S^+)^2 - S^-S^+  - S^+S^- \right].
\end{align}
\end{subequations}
Here the perpendicular operators are changing the $S^\parallel$ quantum number by 0 and $\pm2$, and the $P^{\parallel}$ parallel operator by $\pm1$. We note that
$P^{\parallel}_\text{polar} = P^{\perp 1}_\text{chiral}$
and
$P^{\perp 1}_\text{polar} = -P^{\parallel}_\text{chiral}$,
reflecting the spin-quadrupolar nature of the polarization operators.

As shown in Fig.~\ref{fig:two_mapping}(c), the symmetry group of the spin Hamiltonian is now $C_{2v}\left(C_{1h}\right)=2'm'm$ with elements
\begin{eqnarray}
 C_{2v}\left(C_{1h}\right)=2'm'm=\left\{\mathbf{1}, \sigma^{\parallel},  \Theta C_2^{\perp 2}, \Theta \sigma^{\perp 1} \right\}.
\end{eqnarray}
The character table and the transformations of physical quantities under this group are given in Table \ref{table:chartab_chiral_polar}.

If we compare to the chiral case, firstly, we observe that the $P^{\perp 2}$ transforms as identity, therefore electric polarization along ${\perp 2}$ axis is now allowed by the symmetry hence we refer to this case as the ${\it polar}$ case. 
Secondly, because of the $\sigma^{\parallel}$ symmetry element the $S^{\perp 1}$ and $P^{\perp 2}$ do not belong to the same irreducible representation anymore, and similarly  $S^{\perp 2}$ and $P^{\perp 1}$. 
As a consequence, the NDD in the Faraday geometry vanishes.
Indeed, in Faraday geometry $\kvec/ \| \bdc/$,  the THz field transforms under $\sigma^{\parallel}$ as $\bac/ \to -\bac/$, $\eac/ \to \eac/$, and $\mathbf{k} \to -\mathbf{k} $.
The two directions of the radiation propagation are connected by the  symmetry element and NDD is forbidden.\cite{Szaller2013,Szaller2014}

Instead, the $S^{\perp 2}$ and $P^{\parallel}$ transform according to the same irreducible representation $\mathsf{B_-}$. 
As a consequence, a finite ME susceptibility $\chi^{me}_{\perp2,\|}$ will appear in the Voigt geometry when the $\mathbf{k}$ is parallel to the $\perp\!\!1$ and $\eac/ \parallel \bdc/$. 
Similarly, the $S^{\parallel}$ and $P^{\perp 2}$ both belong to  $\mathsf{A_+}$, providing a finite $\chi^{me}_{\|,\perp2}$. 
Since the operators belonging to the same irreducible representation in Table\,\ref{table:chartab_chiral_polar} have the same parity under $\Theta C_2^{\perp 2}$ transformation, their matrix elements are either both real (for $S^{\perp 2}$ and $P^{\parallel}$), or are both pure imaginary (for  $S^{\parallel}$ and $P^{\perp 2}$), allowing NDD in the Voigt configuration in both polarizations of THz radiation.

As the $\sigma^{\parallel}$ and $C_2^{\parallel}$ are given by the same matrix in Eq.~(\ref{eq:C2parmat}),\cite{altmann1994point} we can repeat the arguments we used in the chiral case to determine the selection rules. 
The $S^{\parallel}$ and $P^{\perp 2}$  operators have finite matrix elements between states with the same $\hat C_2^{\parallel}\equiv \hat \sigma^{\parallel}$ quantum number, i.e. the $\mathcal{A}$-type transitions $1\to 3$ and $2 \to 4$ are allowed.
The  $S^{\perp 2}$ and $P^{\parallel}$ change the quantum number   $\pm i\to\mp i$  and the $\mathcal{B}$-type transitions  $1\to 2$,  $1\to 4$, $2\to 3$, and  $3 \to 4$ are allowed.

\subsubsection{Low symmetry case: $\varphi\neq 0$ nor $\frac{\pi}{4}$}
\label{sec:lowsym}

For arbitrary direction of the external magnetic field within the $XY$ plane, the polarization operators can be written as a linear combination of the chiral and polar cases considered above,
\begin{equation}
\mathbf{P} = \cos 2\varphi \; \mathbf{P}_\text{chiral}
+ \sin 2\varphi  \; \mathbf{P}_\text{polar}.
\end{equation}
Here, only the $\Theta C_2^{\perp 2}$ symmetry remains and the magnetic point group is reduced to the $D_1(C_1)=2'$ (see Table\,\ref{table:chartab_chiral_polar} and Fig.~\ref{fig:two_mapping}(b)).
Since neither $C_2^{\parallel}$ nor $\sigma^{\parallel}$ is a symmetry element of the full problem, 
the \Pvec/ may have finite matrix elements between any of the $(\pm i)$ states, 
and the selection rules we established for the chiral ($\varphi=0$) and the polar ($\varphi=\pi/4$) case are not valid. 
However, as the spin state still respects the quantum number set by the $C_2^{\parallel}$, the associated selection rules, Eq.\,(\ref{eq:AB_selection_rule}),  hold for $S^\parallel$, $S^{\perp 1}$, and  $S^{\perp 2}$. 
Furthermore, since the ME susceptibility is composed from a product of  matrix elements of \Svec/ and \Pvec/, it inherits the selection rules of the  matrix element of \Svec/. 
Putting all this together, the single ion system shows NDD in the Faraday geometry for the  $1\to 2$, $2\to 3$, $1\to 4$, and $3\to 4$ transitions with  $\chi^{me}\propto \cos 2\varphi$ coming from the chiral part, and NDD in the Voigt geometry according to the selection rules set by the polar case with $\chi^{me}\propto \sin 2\varphi$.

\section{The lattice problem}\label{sec:lattice}

This section describes the selection rules when the tetrahedra form a lattice. Furthermore, we give the analytical form  of the transition energies by taking into account the exchange coupling between the spins in the lowest order in perturbation theory. 

\subsection{Magneto-chiral dichroism and selection rules in the lattice problem}

In the material the oxygen tetrahedra are rotated alternatingly, there is no direction of the external field which would show purely either the chiral or the polar case discussed in the section above.
However, the situation is not hopeless: if the external field is along the [100] direction (as in the actual experiment), there is a $\{C_2^\parallel|[\frac{1}{2}00]\}$ screw axis (shown in Fig.~\ref{fig:2sublatt} as $2_1$) which is a symmetry operation. 
The screw axis performs a $\pi$ rotation about the [100] axis and a half translation along the same axis to move the A tetrahedra into B and vice versa (see Fig.~\ref{fig:2sublatt}).  
The screw axis is a symmetry element of the lattice spin Hamiltonian symmetry group, so we can use its irreducible representations to label the eigenstates and  the operators. 
In addition, the $\{\Theta C_2^{\perp 2}|[000]\}$
-- the rotation by $\pi$ about the [001] axes through the center of a tetrahedron followed by a time reversal operation -- is a symmetry element irrespectively from the direction of the in-plane magnetic field.
In fact, this nonsymmorphic magnetic point group is isomorphic to the $D_{2}\left(C_{2}\right)=22'2'$ magnetic point group of a single tetrahedron in the chiral case:
\begin{align}
  \left\{\mathbf{1}, \{C_2^\parallel|[\frac{1}{2}00]\} ,  \{\Theta C_2^{\perp 2}|[000]\},\{\Theta C_2^{\perp 1}|[\frac{1}{2}00]\} \right\}
    \cong\nonumber\\ 
\cong\left\{\mathbf{1}, C_2^{\parallel}, \Theta C_2^{\perp 2}, \Theta C_2^{\perp 1} \right\}.
\end{align}

Let us examine the selection rules based on what we learned for a single ion.
 We express the magnetization,  Eq.~(\ref{eq:Mtot}),  and the polarization,  Eq.~(\ref{eq:Ptot}), in the magnetic-field-fixed coordinate system and decompose into the irreducible representations of the unitary part of the point group, which consists of the identity and the $\{C_2^\parallel|[\frac{1}{2}00]\}$ screw axis. 
The screw axis acts on the magnetization as
\begin{subequations}
\label{eq:screwaxisL}
\begin{align}
\left(\begin{array}{c}
M^\parallel_\mathrm A\\
M^{\perp 1}_\mathrm A\\
M^{\perp 2}_\mathrm A
\end{array}\right)&\rightarrow
\left(\begin{array}{c}
M^\parallel_\mathrm B\\
-M^{\perp 1}_\mathrm B\\
-M^{\perp 2}_\mathrm B
\end{array}\right)
,\\
\left(\begin{array}{c}
M^\parallel_\mathrm B\\
M^{\perp 1}_\mathrm B\\
M^{\perp2}_\mathrm B
\end{array}\right)&\rightarrow
\left(\begin{array}{c}
M^\parallel_\mathrm A\\
-M^{\perp1}_\mathrm A\\
-M^{\perp2}_\mathrm A
\end{array}\right).
\end{align}
\end{subequations}
Similar considerations hold for the polarization operators.

Since the unitary part of the point group has two irreducible representations, the $\mathbf{M}$ operator can be decomposed into even ($ \mathbf{M}_{\boldsymbol{0}} $) and odd ($ \mathbf{M}_{\boldsymbol{\pi}}$) parts as
$\mathbf{M} =  \mathbf{M}_{\boldsymbol{0}} + \mathbf{M}_{\boldsymbol{\pi}}$.

\subsubsection{Selection rules for the even  $(\boldsymbol{0})$ components}
 Using the transformation rules given by Eq.\,(\ref{eq:screwaxisL}) the even part of \Mvec/ is
\begin{subequations}\label{eq:M_even}
\begin{align}
M^\parallel_{\boldsymbol{0}} &= M^\parallel_\mathrm A + M^\parallel_\mathrm B\,,\\
M^{\perp}_{\boldsymbol{0}}&=0,
\end{align}
\end{subequations}
and for the polarizations we get
\begin{subequations}\label{eq:P_even}
\begin{align}
P^\parallel_{\boldsymbol{0}} &=
 \sin 2 \kappa \sum_j  (-1)^j P^{\parallel}_{j,\text{polar}} 
  - \cos 2 \kappa \sum_j P^{\parallel}_{j,\text{chiral}}\,,\\
P^{\perp}_{\boldsymbol{0}}&=0,
\end{align}
\end{subequations}
where $\perp=\bot1, \bot2$ and  the index $j$ runs over all the sites, being an even integer on A  and odd integer on the  B sublattice;
$P^{\parallel}_{j,\text{chiral}}$ is defined by Eq.~(\ref{eq:P_par_chiral}) and $P^{\parallel}_{j,\text{polar}}$ by Eq.~(\ref{eq:P_par_polar}), with the corresponding spin operators at site $j$. 
The light  does not interact with the system in the Faraday geometry in the even channel because $P^{\perp}_{\boldsymbol{0}}= M^{\perp}_{\boldsymbol{0}} =0$.

\subsubsection{Selection rules for the odd  $(\boldsymbol{\pi})$ components}
\label{sec:selectionpi}
When we antisymmetrize, we get the odd quantities  $\mathbf{M}_{\boldsymbol{\pi}} $ and $\mathbf{P}_{\boldsymbol{\pi}} $, and following the steps we used to obtain the even components above, the corresponding magnetization and polarization components are
\begin{subequations}
\begin{align}
M^\parallel_{\boldsymbol{\pi}} &=0,\\
M^{\perp}_{\boldsymbol{\pi}}&= M^{\perp}_\mathrm A + M^{\perp}_\mathrm B  
\end{align}
\end{subequations}
and
\begin{subequations}
\begin{align}
P^\parallel_{\boldsymbol{\pi}} &=0,\\
P^{\perp}_{\boldsymbol{\pi}} &=
 \sin 2 \kappa \sum_j  (-1)^j P^{\perp}_{j,\text{polar}}  
- \cos 2 \kappa \sum_j P^{\perp}_{j,\text{chiral}}\,.
\end{align}
\end{subequations}
The non-vanishing perpendicular components $\{M^{\perp1}_{\boldsymbol{\pi}},P^{\perp2}_{\boldsymbol{\pi}}\}$ and $\{M^{\perp2}_{\boldsymbol{\pi}}, P^{\perp1}_{\boldsymbol{\pi}}\}$ lead to a finite ME susceptibility and NDD in the Faraday geometry for the odd channel.

We now examine the effect of the time reversal. The action of the $\{\Theta C_2^{\perp 2}|[000]\}$ is given by
\begin{subequations}
\label{eq:ThetaC2vL}
\begin{align}
\left(\begin{array}{c}
M^\parallel_\mathrm A\\
M^{\perp 1}_\mathrm A\\
M^{\perp 2}_\mathrm A
\end{array}\right)&\rightarrow
\left(\begin{array}{c}
M^\parallel_\mathrm A\\
M^{\perp 1}_\mathrm A\\
-M^{\perp 2}_\mathrm A
\end{array}\right),\\
\left(\begin{array}{c}
P^\parallel_\mathrm A\\
P^{\perp 1}_\mathrm A\\
P^{\perp2}_\mathrm A
\end{array}\right)&\rightarrow
\left(\begin{array}{c}
-P^\parallel_\mathrm A\\
-P^{\perp1}_\mathrm A\\
P^{\perp2}_\mathrm A
\end{array}\right),
\end{align}
\end{subequations}
and the same equations on the B sublattice. 
Just like in the single ion problem, the $M^{\perp1}_{\boldsymbol{\pi}}$ and $P^{\perp2}_{\boldsymbol{\pi}}$ belong to the same irreducible representation, as well as the $M^{\perp2}_{\boldsymbol{\pi}}$ and $P^{\perp1}_{\boldsymbol{\pi}}$. 
The matrix elements are therefore real or imaginary, and the product of the magnetization and polarization matrix elements in the ME susceptibility is  real.

Although the symmetry classification obtained above did not consider  the DM interaction, it describes the selection rules obtained from the exact diagonalization. 
This is  because the DM interaction   is compatible with the $D_{2}\left(C_{2}\right)$ magnetic point group considered above.

\subsection{Perturbative effects of the exchange coupling}

So far, we have neglected the interactions between the Co spins. To assess how much the single ion picture is altered, we take into account the effect of exchange couplings perturbatively: starting from the single-ion limit, $J=J_{z}=D_z=0$ in Eq.~(\ref{eq:Hamiltonian}), we derive a tight-binding-like approximation for the exchange part. 
 We denote the on-site and exchange parts as $\mathcal{H}^0$ and $\mathcal{H}'$.

For the noninteracting case, the ground state of $\mathcal{H}^0$ is 
\begin{align}
 | \Psi_1^{0}\rangle = 
 \prod_{j} |\psi_1^{(-i)}(j)\rangle  \,,
 \end{align}
 where $j$ runs over both the A and B sublattice sites.
 We  define the local single-ion excitation at site $l$ as 
 \begin{align}
 | \Psi_{2}^{0} (l)\rangle &= 
  | \psi_{2}^{(+i)}(l)\rangle \prod_{j \neq l} | \psi_{1}^{(-i)}(j)\rangle \;.
 \end{align}
The local wavefunctions $| \psi_{\alpha}^{(\pm i)}(l)\rangle $ are given by  Eq.\,(\ref{eq:eigen}). 
If the number of (all) sites is $N$, the noninteracting ground state energy is $E_{1}^{0}=N\varepsilon_1$ and the excited states  $|\Psi_{2}^{0} (l)\rangle $ have energies $E_{2}^{0}=(N-1)\varepsilon_1+\varepsilon_2$ and are $N$-fold degenerate. 

In order to calculate the degeneracy lifting in the first order of the degenerate perturbation expansion, we need to diagonalize the perturbing matrix $\mathcal{H}'$ on the subspace spanned by $| \Psi_{2}^0 (l)\rangle$, 
\begin{eqnarray}
\left\langle \Psi_{2}^0 (l') \right| \mathcal{H}' \left|\Psi_{2}^0 (l)\right\rangle,
\end{eqnarray}
describing a local excited state which hops with equal amplitudes in different directions. 
Generally, the hopping problem is diagonalized in the momentum space for a translationally symmetric problem. 
In our case, the translational symmetry holds for the unit cell containing two lattice sites, with translation vectors $\mathbf{t}'_1=(1,1,0)$ and $\mathbf{t}'_2=(1,-1,0)$ in the $(x,y,z)$ coordinate system (see Fig.~\ref{fig:2sublatt}). 
However, the $\{C_2^\parallel|[\frac{1}{2}00]\}$ screw axis and the $\mathbf{t}'_1$ and/or $\mathbf{t}'_2$  translation generate an abelian group isomorphic to the group constructed from the $\mathbf{t}_1=(1,0,0)$ and $\mathbf{t}_2=(0,1,0)$ -- the translations of the lattice of the Co ions, if we neglect the alternating tetrahedra. 
The irreducible representations are all one dimensional, with
\begin{align}
 \{C_2^\parallel|[\frac{1}{2}00]\} |\Psi_\mathbf{k} \rangle &= e^{i \mathbf{k}\cdot\mathbf{t}_1} |\Psi_\mathbf{k} \rangle \;,
 \end{align}
where $\mathbf{k}$ plays the role of the momentum. 
As a result, the $\mathbf{k}=(0,0)=\boldsymbol{0}$ and $\mathbf{k}=(\pi,\pi)=\boldsymbol{\pi}$ states (which,  in fact, are both at the $\Gamma$ point of the Brillouin zone of the lattice defined by the proper translations $\mathbf{t}'_1$ and $\mathbf{t}'_2$) are realized by the following linear combinations:
\begin{align}
 | \Psi_{2,\boldsymbol{0}}^0 \rangle &= 
 \frac{1}{\sqrt{N}}\left(
 \sum_{j \in A}  | \Psi_{2}^0  (l)\rangle + C_2^\parallel \sum_{j\in B}  | \Psi_{2}^0  (l)\rangle \right), \nonumber\\
 | \Psi_{2,\boldsymbol{\pi}}^0 \rangle &= 
\frac{1}{\sqrt{N}}\left(
 \sum_{j \in A}  | \Psi_{2}^0  (l)\rangle - C_2^\parallel \sum_{j\in B}  | \Psi_{2}^0  (l)\rangle \right).
\end{align}
Then the energies of the excitations are 
\begin{align}
  \omega_{\mathbf{k}}^{1\to2} &= E^{(1)}_{\mathbf{k},2} - E^{(1)} \nonumber\\
   &= \langle \Psi_{2,\mathbf{k}}^0| \mathcal{H} | \Psi_{2,\mathbf{k}}^0 \rangle - \langle \Psi_{1}^0 | \mathcal{H} | \Psi_{1}^0 \rangle \,.
\end{align}
 The evaluation of the expectation values using the single ion wave functions Eq.~(\ref{eq:eigen}) is straightforward but tedious. 

Keeping only the first-order terms of the perturbation and doing series expansion  we get in the  strong magnetic field limit, $h \gg \Lambda \gg J,J_z$, and $D_z=0$,
\begin{subequations}
\begin{align}
\omega^{1\to 2}_{\boldsymbol{0}} &=  h  + \Lambda  - 9 J  - 3 J_z,  \\
 \omega^{1\to 2}_{\boldsymbol{\pi}} &= h  + \Lambda  - 3 J  + 3 J_z. \label{eq:w12pi}
 \end{align}
\end{subequations}
The perturbative approach works well for the experimentally studied paramagnetic case in the high-field limit. In Sec. \ref{sec:selectionpi} we have seen that only the $\boldsymbol{\pi}$ modes absorb in Faraday geometry. 
In the exact diagonalization, we observed the strongest absorption is for the  $\omega^{1\to 2}_{\boldsymbol{\pi}}$ mode. 
This is not surprising, as this is the  single-magnon  mode in the weak anisotropy limit, $\Lambda\ll J, J_z$.  
In the single-ion model without anisotropy, $J,J_z,\Lambda=0$, it is just the paramagnetic mode with $\omega^{1\to 2}_{\boldsymbol{\pi}} =h$, as it can be inferred from Eq.~(\ref{eq:w12pi}). 
More interestingly, the energy of this mode is very close to the single-ion excitation energy $\omega^{1\to 2} =\varepsilon_2-\varepsilon_1$, see Eq.~(\ref{eq:varepsilons_bigh}).
The difference is just $3(J  - J_z)$ and vanishes for $J=J_z$. 
This explains why the single-ion model works remarkably well for large magnetic fields, $h \gg \Lambda,J,J_z$, as seen from the comparison of the minimal theory and the experiments in Fig.~\ref{fig:single_ion_e_de}.

In  the strong single-ion anisotropy  limit, $\Lambda \gg h  , J,J_z$, the energies of these modes are
\begin{subequations}
	\begin{align}
	\omega^{1\to 2}_{\boldsymbol{0}} &= 2 h  - 12 J  - J_z, \label{eq:w120_L}\\
	\omega^{1\to 2}_{\boldsymbol{\pi}} &= 2 h  - 4 J  + J_z.
	\end{align}
\end{subequations}
The deviation of $\omega^{1\to 2}_{\boldsymbol{\pi}}$ in the single-ion model becomes noticeable compared to the strong field limit given by Eq.~(\ref{eq:w12pi}) , $-4J+J_z$ vs. $-3J+3J_z$,  but still not too large as compared to the deviations of other modes, say Eq.~(\ref{eq:w120_L}).

\section{Summary}\label{sec:summary}

 The coupling of the magnetic moments and electric polarization in a material is responsible for many interesting optical phenomena which happen at low temperatures, usually in the ordered phases. Here we studied the optical response of a multiferroic material in the paramagnetic phase, at temperatures much higher than the ordering temperature.   
   
 We observed MChD  for the excitations of the $S=3/2$ spin of the  Co$^{2+}$ ion in \scso/ at temperatures up to 100\,K, although   the magnetic ordering temperature is only $T_\mathrm{N}=7$\,K. 
The main experimental findings -- the temperature dependence of the spin mode frequencies, their intensities, and the sign of the MChD for the different spin modes -- are captured well by the exact diagonalization of a  four-spin cluster. 

The numerical results are interpreted in a simple analytical model  of a single spin with an easy-plane anisotropy in an external magnetic field,  and its coupling to the THz radiation via magnetization \Mvec/ and via polarization \Pvec/, expressed by  spin-quadrupolar terms. Finite ME susceptibility arises if the components of the \Mvec/ and \Pvec/ transform according to the same irreducible representation of the symmetry group compatible with both, the oxygen tetrahedron around the magnetic ion and the applied magnetic field. Fig.~\ref{fig:single_ion_e_de} shows that in high fields and low temperatures ($\varepsilon_2-\varepsilon_1 > k_\mathrm BT$) the NDD spectrum is dominated by the magnetic dipolar transition from the ground state, $\ket{1}\rightarrow \ket{2}$. By increasing $T$, another magnetic dipolar transition from the thermally excited state, $\ket{2}\rightarrow \ket{3}$, appears in the NDD spectrum. 
Our exact diagonalization calculation showed that the $\ket{2}\rightarrow \ket{3}$ peak is very close to the $\ket{1}\rightarrow \ket{2}$ in energy 
and due to the line-broadening effects only a single broad peak is observed. 
This coincides with the experimental finding at high temperatures, as exemplified in Fig.~\ref{fig:scso_b[100]_14t_experiment}\,(a) and (b). 
 
By considering the real material where the  the oxygen tetrahedra are tilted, we showed how the selection rules of the single-ion model are modified when the exchange coupling is turned on, in agreement with the results of the exact diagonalization. 
Finally, we demonstrated that the exchange  correlations are important to accurately describe the mode frequencies in the paramagnetic state. 

In conclusion, we demonstrated that MChD can arise in the paramagnetic phase of a non-centrosymmetric material. Furthermore, we presented a detailed theoretical analysis of spin excitations in \scso/ which helps to identify the key parameters responsible for high temperature NDD both in the chiral and toroidal cases.

\begin{acknowledgments}
This research was supported by the by the Estonian Ministry of Education and Research Grant IUT23-3, 
by the European Regional Development Fund project TK134, 
by the bilateral program of the Estonian and Hungarian Academies of Sciences under the Contract No. SNK-64/2013, 
by the 
Hungarian NKFIH Grants No. K 124176 and ANN 122879, 
by the BME-Nanonotechnology and Materials Science FIKP grant of EMMI (BME FIKP-NAT),
by the FWF Austrian Science Fund I 2816-N27, and by the Deutsche Forschungsgemeinschaft
(DFG) via the Transregional Research
Collaboration TRR 80: From Electronic Correlations to
Functionality (Augsburg—Munich—Stuttgart)
We acknowledge the support of the HFML-RU/FOM, member of the European Magnetic Field Laboratory (EMFL).
\end{acknowledgments}

\section*{Author Contributions}
V.K. and Y. Tokunaga grew the crystals, 
F.D., D.Sz., V.K., J.V. and U.N.  conducted THz spectroscopy experiments in Tallinn and analyzed the results,
F.D., D.Sz., B.B. and D.L.K. conducted high magnetic field measurement in Nijmegen, 
P.B. and K.P. developed the theory, 
T.R., S.B., P.B., and K.P. wrote the manuscript,
D.Sz. and K.P. conceived the project.
Every author contributed to the discussion of the results.

\appendix

\section{Reality of the matrix elements in a magnetic point group}
\label{sec_TBD}
We analyze the case of $\varphi=0$, which is relevant for the experimental situation and is conceptually the simplest. The symmetry group of this configuration considering both the spin Hamiltonian, Eq.\,(\ref{eq:H_1ion}), and the oxygen tetrahedron around the Co$^{2+}$ ion, manifested by the spin-polarization coupling (\ref{eq:pop1}), is  
\begin{eqnarray}
D_{2}\left(C_{2}\right)=\left\{\mathbf{1}, C_2^{\parallel},  \Theta C_2^{\perp 1}, \Theta C_2^{\perp 2} \right\},
\end{eqnarray}
with operators in the field-fixed coordinate frame, Fig.\,\ref{fig:3Dcoords}. 
Since the inversion, the mirror plane, and the roto-reflection symmetry are absent, the case is termed as  chiral. 
The $D_{2}\left(C_{2}\right)$ group is generated by the symmetry elements $C_2^{\parallel}$ and $\Theta C_2^{\perp 2}$, and its character table together with the symmetry classification of spin and polarization components is summarized in Table\,\ref{table:chartab_chiral_polar}. 
We note that our conclusions are equally valid for the $\varphi=\pi/4$ case, where $\Theta C_2^{\perp 2}$ is also present in the corresponding polar symmetry group $C_{2v}\left(C_{1h}\right)$.

The operator $C_2^{\perp 2}$ is represented on the $S=3/2$ spin space by the matrix
\begin{equation}
\hat{C}_2^{\perp 2}=e^{i \pi \hat{S}^{\perp 2}}=
\left(
\begin{array}{cccc}
0& 0 & 0 & 1 \\
0 & 0 & -1 & 0 \\
0 & 1 & 0 & 0 \\
-1 & 0 & 0 & 0 \\
\end{array}
\right).
\label{eq:C2perp2mat}
\end{equation}
The time-reversal operation $\Theta$ is conveniently represented by the antiunitary operation \begin{equation}
\hat{\Theta}=\hat{C}_2^{\perp 2}\mathcal{K}=e^{i \pi \hat{S}^{\perp 2}}\mathcal{K}=
\left(
\begin{array}{cccc}
0& 0 & 0 & 1 \\
0 & 0 & -1 & 0 \\
0 & 1 & 0 & 0 \\
-1 & 0 & 0 & 0 \\
\end{array}
\right)\mathcal{K},
\label{eq:thetamat}
\end{equation}
where $\mathcal{K}$ is the complex conjugation, conjugating every matrix or vector component on the right, but leaving the basis functions intact.\cite{sakurai} This is analogous to the $S=1/2$ case where $\Theta$ is expressed with the help of the Pauli matrix $\hat\sigma^y$ as $\hat{\Theta}=i\hat{\sigma}^y\mathcal{K}$.
Collecting all this together, the operator for $\Theta C_2^{\perp 2}$ in the Hilbert-space reads as
\begin{equation}
\hat{\Theta}\hat{C}_2^{\perp 2}=\hat{C}_2^{\perp 2}\mathcal{K}\hat{C}_2^{\perp 2}=e^{i 2\pi \hat{S}^{\perp 2}}\mathcal{K}=-\hat{\mathbf{1}}\mathcal{K}.
\label{eq:thetaC2perp2mat}
\end{equation}
When we consider the effect of this symmetry on the matrix elements of physical observables, we find firstly, all representations of $\Theta C_2^{\perp 2}$ are $+1$ or $-1$ as given in table \ref{table:chartab_chiral_polar}  for the chiral and polar cases, respectively. 
Therefore, any linear operator $\mathcal{O}$ -- such as the spin or polarization components -- is either even or odd, i.e. it transforms as
\begin{eqnarray}
\Theta C_2^{\perp 2}\left(\mathcal{O}\right) =\pm \mathcal{O},
\end{eqnarray}
or in matrix representation:
\begin{eqnarray}
\left(\hat{\Theta}\hat{C}_2^{\perp 2} \right)\hat{\mathcal{O}} \left(\hat{\Theta}\hat{C}_2^{\perp 2}\right)^{-1}=\pm \hat{\mathcal{O}}.
\label{signAperp}
\end{eqnarray}
Next, following Eq.~(\ref{eq:thetaC2perp2mat}), let us evaluate the left hand side of the equation above: \begin{eqnarray}
\left(\hat{\Theta}\hat{C}_2^{\perp 2}\right) \hat{\mathcal{O}}\left(\hat{\Theta}\hat{C}_2^{\perp 2} \right)^{-1}=\mathcal{K}\hat{\mathcal{O}}\mathcal{K}=\hat{\mathcal{O}}^*,
\label{conjAperp}
\end{eqnarray}
where the last equality follows from the fact that acting on an arbitrary vector $v$,
$
\mathcal{K}\hat{\mathcal{O}}\mathcal{K} v=\mathcal{K}\hat{\mathcal{O}} v^*=\hat{\mathcal{O}}^* v.
$
Eqs.~(\ref{signAperp}) and (\ref{conjAperp}) together mean that
\begin{eqnarray}
\hat{\mathcal{O}}^* =\pm \hat{\mathcal{O}}.
\end{eqnarray}
This restricts the matrix elements of the operators even under $\Theta C_2^{\perp 2}$ to be real, and the matrix elements of the odd operators to be pure imaginary.

In conclusion, the unitary symmetries determine the selection rules i.e.~the non-vanishing matrix elements, whereas antiunitary elements give constraints on the reality of them. These rules and the matrix elements given in Appendix~\ref{sec_matrix_elements} (and calculated by using the explicit wavefunctions) are in full agreement.

\section{Matrix elements in the single-ion problem}
\label{sec_matrix_elements}
\subsection{Large anisotropy and small field: $\Lambda \gg h$}

From the series expansion of the single spin energies (\ref{eq:1ion_energies}) in $h/\Lambda$, we get
\begin{subequations}
\label{eq:varepsilons_bigLambda}
\begin{align}
\varepsilon_1 &= \frac{\Lambda }{4} -h-\frac{3 h^2}{8 \Lambda } + \cdots, \\
\varepsilon_2 &= \frac{\Lambda }{4}+ h-\frac{3 h^2}{8 \Lambda } + \cdots, \\
\varepsilon_3 &= \frac{9 \Lambda }{4}  +\frac{3 h^2}{8 \Lambda } + \cdots,\\
\varepsilon_4 &= \frac{9 \Lambda }{4} +\frac{3 h^2}{8 \Lambda } + \cdots.
\end{align}
\end{subequations}
The spin  and polarization operators in  leading order of $h/\Lambda$ are in the matrix representation
\begin{subequations}
\begin{align}
\hat{S}^{\|} &=\left(
\begin{array}{cccc}
 1 & 0 & -\frac{\sqrt{3}}{2} & 0 \\
 0 & -1 & 0 & -\frac{\sqrt{3}}{2} \\
 -\frac{\sqrt{3}}{2} & 0 & \frac{9 h^2}{16 \Lambda ^2} & 0 \\
 0 & -\frac{\sqrt{3}}{2} & 0 & -\frac{9 h^2}{16 \Lambda ^2} \\
\end{array}
\right),
\\
\hat{S}^{\perp 1} &=\left(
\begin{array}{cccc}
 0 & 1 & 0 & -\frac{\sqrt{3}}{2} \\
 1 & 0 & \frac{\sqrt{3}}{2} & 0 \\
 0 & \frac{\sqrt{3}}{2} & 0 & \frac{9 h^2}{16 \Lambda ^2} \\
 -\frac{\sqrt{3}}{2} & 0 & \frac{9 h^2}{16 \Lambda ^2} & 0 \\
\end{array}
\right),
\\
\hat{S}^{\perp 2} &= i\left(
\begin{array}{cccc}
 0 & -\frac{1}{2} & 0 & \frac{\sqrt{3} h}{4 \Lambda } \\
 \frac{1}{2} & 0 & -\frac{\sqrt{3} h}{4 \Lambda } & 0 \\
 0 & \frac{\sqrt{3} h}{4 \Lambda } & 0 & -\frac{3}{2} \\
 -\frac{\sqrt{3} h}{4 \Lambda } & 0 & \frac{3}{2} & 0 \\
\end{array}
\right),
\end{align}
\end{subequations}
\begin{subequations}
\begin{align}
\hat{P}^{\|}_{\text{chiral}} &= - \hat{P}^{\perp 1}_{\text{polar}} =i\left(
\begin{array}{cccc}
	   0     &    0     & - \sqrt{3} &     0      \\
	   0     &    0     &     0      & - \sqrt{3} \\
	\sqrt{3} &    0     &     0      &     0      \\
	   0     & \sqrt{3} &     0      &     0
\end{array}
\right),
\\
\hat{P}^{\|}_{\text{polar}} &= \hat{P}^{\perp 1}_{\text{chiral}} =
i\left(
\begin{array}{cccc}
	           0            & -\frac{3  h}{2 \Lambda } &            0             &        \sqrt{3}         \\
	\frac{3  h}{2 \Lambda } &            0             &        - \sqrt{3}        &            0            \\
	           0            &         \sqrt{3}         &            0             & \frac{3  h}{2 \Lambda } \\
	      - \sqrt{3}        &            0             & -\frac{3  h}{2 \Lambda } &            0
\end{array}
\right),
\\
\hat{P}^{\perp 2}_{\text{chiral}} &=
\left(
\begin{array}{cccc}
 0 & \frac{3 h}{2 \Lambda } & 0 & -\sqrt{3} \\
 \frac{3 h}{2 \Lambda } & 0 & -\sqrt{3} & 0 \\
 0 & -\sqrt{3} & 0 & -\frac{3 h}{2 \Lambda } \\
 -\sqrt{3} & 0 & -\frac{3 h}{2 \Lambda } & 0 \\
\end{array}
\right),
\\
\hat{P}^{\perp 2}_{\text{polar}} &=
\left(
\begin{array}{cccc}
 \frac{3 h}{2 \Lambda } & 0 & -\sqrt{3} & 0 \\
 0 & -\frac{3 h}{2 \Lambda } & 0 & \sqrt{3} \\
 -\sqrt{3} & 0 & -\frac{3 h}{2 \Lambda } & 0 \\
 0 & \sqrt{3} & 0 & \frac{3 h}{2 \Lambda } \\
\end{array}
\right).
\end{align}
\end{subequations}

\subsection{Large field and small anisotropy: $h \gg \Lambda $}

The energies are
\begin{subequations}
\label{eq:varepsilons_bigh}
\begin{align}
\varepsilon_1 &=  -\frac{3 h}{2}+\frac{3 \Lambda }{4}-\frac{3 \Lambda ^2}{8 h} + \cdots, \\
\varepsilon_2 &=  -\frac{h}{2}+\frac{7 \Lambda }{4}-\frac{3 \Lambda ^2}{8 h} + \cdots, \\
\varepsilon_3 &=  \frac{h}{2}+\frac{7 \Lambda }{4}+\frac{3 \Lambda ^2}{8 h} + \cdots, \\
\varepsilon_4 &=  \frac{3 h}{2}+\frac{3 \Lambda }{4}+\frac{3 \Lambda ^2}{8 h} + \cdots.
\end{align}
\end{subequations}
The spin  and polarization operators in the leading order of $\Lambda/h$ are in the matrix representation
\begin{subequations}
\begin{align}
\hat{S}^{\|} &= \left(
\begin{array}{cccc}
 \frac{3}{2} & 0 & -\frac{\sqrt{3} \Lambda }{2 h} & 0 \\
 0 & \frac{1}{2} & 0 & -\frac{\sqrt{3} \Lambda }{2 h} \\
 -\frac{\sqrt{3} \Lambda }{2 h} & 0 & -\frac{1}{2} & 0 \\
 0 & -\frac{\sqrt{3} \Lambda }{2 h} & 0 & -\frac{3}{2} \\
\end{array}
\right),
\\
\hat{S}^{\perp 1} &= \left(
\begin{array}{cccc}
 0 & \frac{\sqrt{3}}{2} & 0 & -\frac{9 \Lambda ^2}{16 h^2} \\
 \frac{\sqrt{3}}{2} & 0 & 1 & 0 \\
 0 & 1 & 0 & \frac{\sqrt{3}}{2} \\
 -\frac{9 \Lambda ^2}{16 h^2} & 0 & \frac{\sqrt{3}}{2} & 0 \\
\end{array}
\right),
\\
\hat{S}^{\perp 2} &=i\left(
\begin{array}{cccc}
 0 & -\frac{\sqrt{3}}{2} & 0 & \frac{3  \Lambda ^2}{16 h^2} \\
 \frac{ \sqrt{3}}{2} & 0 & -1 & 0 \\
 0 & 1 & 0 & -\frac{\sqrt{3}}{2}  \\
 -\frac{3  \Lambda ^2}{16 h^2} & 0 & \frac{ \sqrt{3}}{2} & 0 \\
\end{array}
\right),
\end{align}
\end{subequations}
\begin{subequations}
\begin{align}
\hat{P}^{\|}_{\text{chiral}} &= - \hat{P}^{\perp 1}_{\text{polar}} =
i\left(
\begin{array}{cccc}
 0 & 0 & - \sqrt{3} & 0 \\
 0 & 0 & 0 & - \sqrt{3} \\
  \sqrt{3} & 0 & 0 & 0 \\
 0 &  \sqrt{3} & 0 & 0 \\
\end{array}
\right),
\\
\hat{P}^{\|}_{\text{polar}} &= \hat{P}^{\perp 1}_{\text{chiral}} =
i\left(
\begin{array}{cccc}
 0 & - \sqrt{3} & 0 & \frac{3  \Lambda }{2 h} \\
  \sqrt{3} & 0 & -\frac{3  \Lambda }{2 h} & 0 \\
 0 & \frac{3  \Lambda }{2 h} & 0 &  \sqrt{3} \\
 -\frac{3  \Lambda }{2 h} & 0 & - \sqrt{3} & 0 \\
\end{array}
\right),
\\
\hat{P}^{\perp 2}_{\text{chiral}} &=
\left(
\begin{array}{cccc}
 0 & \sqrt{3} & 0 & -\frac{3 \Lambda }{2 h} \\
 \sqrt{3} & 0 & -\frac{3 \Lambda }{2 h} & 0 \\
 0 & -\frac{3 \Lambda }{2 h} & 0 & -\sqrt{3} \\
 -\frac{3 \Lambda }{2 h} & 0 & -\sqrt{3} & 0 \\
\end{array}
\right),
\\
\hat{P}^{\perp 2}_{\text{polar}} &=
\left(
\begin{array}{cccc}
 \frac{3}{2} & 0 & -\frac{\sqrt{3}}{2} & 0 \\
 0 & -\frac{3}{2} & 0 & -\frac{\sqrt{3}}{2} \\
 -\frac{\sqrt{3}}{2} & 0 & -\frac{3}{2} & 0 \\
 0 & -\frac{\sqrt{3}}{2} & 0 & \frac{3}{2} \\
\end{array}
\right).
\end{align}
\end{subequations}

\bibliographystyle{apsrev4-1}
\end{document}